\documentclass[useAMS,usenatbib]{mn2e}
\usepackage{graphicx}
\usepackage{amsmath}
\usepackage{amssymb}
\usepackage{deluxetable}

\newcommand\aj{{AJ}}%
\newcommand\araa{{ARA\&A}}%
\newcommand\apj{{ApJ}}%
\newcommand\apjs{{ApJS}}%
\newcommand\aap{{A\&A}}%
\newcommand\mnras{{MNRAS}}%
\newcommand\pasp{{PASP}}%
\newcommand\iaucirc{{IAU~Circ.}}%
\def\cb{SN\,2000cb} 

\newcommand{\Nifs}{\ensuremath{^{56}\mathrm{Ni}}}
\newcommand{\Msun}{{\ensuremath{\mathrm{M}_{\odot}}}}

\title[Peculiar Type II Supernovae from Blue Supergiants]{Peculiar Type II Supernovae from Blue Supergiants}

\author[Kleiser et al.]
{Io K. W. Kleiser$^{1}$\thanks{E-mail:io.kleiser@berkeley.edu},
Dovi Poznanski$^{2,1,3}$\thanks{E-mail:dovi@berkeley.edu},
Daniel Kasen$^{4,2}$,
Timothy R. Young$^{5}$,\newauthor
 Ryan Chornock$^{6}$, 
 Alexei V. Filippenko$^{1}$,
 Peter Challis$^{6}$,
 Mohan Ganeshalingam$^{1}$,\newauthor
 Robert P. Kirshner$^{6}$, 
 Weidong Li$^{1}$,
 Thomas Matheson$^{7}$,
 Peter E. Nugent$^{2}$,\newauthor
 and Jeffrey M. Silverman$^{1}$\\
\\
$^{1}$Department of Astronomy, University of California,
  Berkeley, CA 94720-3411.\\
$^{2}$Lawrence Berkeley National Laboratory, 1 Cyclotron
  Road, Berkeley, CA 94720.\\
$^{3}$Einstein Fellow.\\
$^{4}$Department of Physics, University of California,
  Berkeley, CA 94720.\\
$^{5}$Department of Physics and Astrophysics, University of North Dakota,
  213 Witmer Hall, Grand Forks, ND 58202-7129.\\
$^{6}$Harvard-Smithsonian Center for Astrophysics, 60
  Garden Street, Cambridge, MA 02138.\\
$^{7}$National Optical Astronomy Observatory, 950 North
  Cherry Avenue, Tucson, AZ 85719.\\}
\begin{document}
\maketitle
\label{firstpage}
\begin{abstract}

The vast majority of Type II supernovae (SNe) are produced by red
supergiants (RSGs), but SN\,1987A revealed that blue supergiants
(BSGs) can produce members of this class as well, albeit with some
peculiar properties. This best studied event revolutionised our
understanding of SNe, and linking it to the bulk of Type II events is
essential.  We present here optical photometry and spectroscopy
gathered for SN\,2000cb, which is clearly not a standard Type II SN
and yet is not a SN\,1987A analog. The light curve of \cb\ is
reminiscent of that of SN\,1987A in shape, with a slow rise to a late
optical peak, but on substantially different time
scales. Spectroscopically, SN\,2000cb resembles a normal SN\,II, but
with ejecta velocities that far exceed those measured for SN\,1987A or
normal SNe\,II, above 18,000\,km\,s$^{-1}$ for H$\alpha$ at early
times.  The red colours, high velocities, late photometric peak, and
our modeling of this object all point toward a scenario involving the
high-energy explosion of a small-radius star, most likely a BSG,
producing 0.1\,$\Msun$ of $^{56}$Ni.  Adding a similar object to the
sample, SN\,2005ci, we derive a rate of $\sim 2$\% of the core-collapse
rate for this loosely defined class of BSG explosions.

\end{abstract}


\section{Introduction}

Massive stars that retain their hydrogen envelope during their
evolution end their lives as Type II supernovae (SNe\,II). The vast
majority of these are characterised by a fast (few days) rise to a
flat light curve, most pronounced in the reddest optical bands, with a
duration of 80--100\,d. This ``plateau'' phase, for which they have
been named SNe\,II-P, is interpreted as the recession of the
photosphere as the ejecta expand and cool
\citep[e.g.,][]{kirshner73, barbon79}. The spectra of SNe\,II-P are typically
dominated by strong P-Cygni profiles of hydrogen lines, as well as
iron absorption features \citep[e.g., see the review
  by][]{filippenko97}.

Due to the relatively simple physics driving their\citet{} optical
evolution, SNe II-P were the first SNe to be suggested as distance
indicators, via the so-called expanding photosphere method (EPM;
\citealt{kirshner74}). While this method and its descendants depend on
modeling, \citet{hamuy02} suggested an empirical relation between
luminosity and ejecta velocity, as measured from iron absorption
lines, that proved compelling when combined with dust-extinction
correction \citep{nugent06,olivares10,poznanski09,poznanski10}.

Over the past decade, $\sim 20$ pre-explosion locations of SNe\,II-P
have been directly imaged with the {\it Hubble Space Telescope} or
deep ground-based images, yielding five detections of progenitor
stars, all of which were red supergiants (RSGs), and many limits on
stars with masses in the range 7.5--15 $\Msun$ \citep[][ and
  references therein]{smartt09}. These masses are somewhat at odds
with those derived from explosion modeling, which tend to be higher,
closer to 15--25 $\Msun$
\citep[e.g.,][]{nadyozhin03,utrobin07,utrobin09}, though
\citet{dessart10b} push toward lower masses of less than $\sim
20$\,$\Msun$.

SNe\,II-P show some diversity, but overall they tend to be a fairly
homogeneous group. However, there are intriguing relatives to this
class of objects, most notably SN\,1987A, which exploded in the nearby
Large Magellanic Cloud and whose progenitor was a compact blue
supergiant (BSG) star \citep[][ and references
  therein]{arnett89}. Given the unique contribution of SN\,1987A to
the study of core-collapse SNe, particular attention to objects that
resemble it is warranted.  The most similar published SN was SN\,1998A
\citep{pastorello05}. While its light-curve shape was nearly identical
to that of SN\,1987A, SN\,1998A was more luminous and bluer, and its
spectra showed higher expansion velocities. \citet{pastorello05}
attribute these differences to a higher-energy BSG explosion.

\cb\ was discovered on 27.4 April 2000 (UT dates are used throughout
this paper) by \citet{papenkova00} in the spiral galaxy IC\,1158 at
$\alpha = 16^{\rm h}01^{\rm m}32.15^{\rm s}$, $\delta =
+1^\circ42\arcmin23\arcsec.0$ (J2000) as part of the Lick Observatory
Supernova Search \citep{li00,filippenko01}. An unfiltered image
obtained with the Katzman Automatic Imaging Telescope (KAIT) on 24.4
April shows the supernova, while there is no detection to a limiting
magnitude of 18.7 in an image taken on 9.5 April. The SN is at a
projected distance of about 4.3\,kpc from the host-galaxy centre, off
one of the outer arms of this spiral galaxy. \citet{jha00} classified
it spectroscopically as a SN\,II on 28.4 April and noted high expansion
velocities measured from the hydrogen absorption features (up to
18300 km\,s$^{-1}$ for H$\alpha$), as confirmed on 29.3 April by
\citet{aldering00}. Some peculiarities of \cb\ and its similarity to
SN\,1987A were noted by \citet{hamuy01b}, and as a result, both SNe
were excluded from a sample of SNe\,II-P used as standardizable candles
\citep{hamuy04}.

In this paper, we present photometric and spectroscopic data on
SN\,2000cb, all taken within 160~d after explosion. We investigate its
properties and compare it to the prototypical Type II SN\,1999em, to
SN\,1987A, and to SN\,1998A. As we show, while \cb\ shares some
characteristics with SNe\,1987A and 1998A, it does not appear to
closely match any of our comparison objects, further expanding the
range of known possible outcomes of massive stellar death. Various
arguments, in addition to our model fit to the bolometric evolution,
point to a BSG progenitor that produced a strong explosion with a
small envelope and a significant amount of $^{56}$Ni.

\section{Data Acquisition and Reduction}

\begin{figure}
\centering
\includegraphics[width=3.2in]{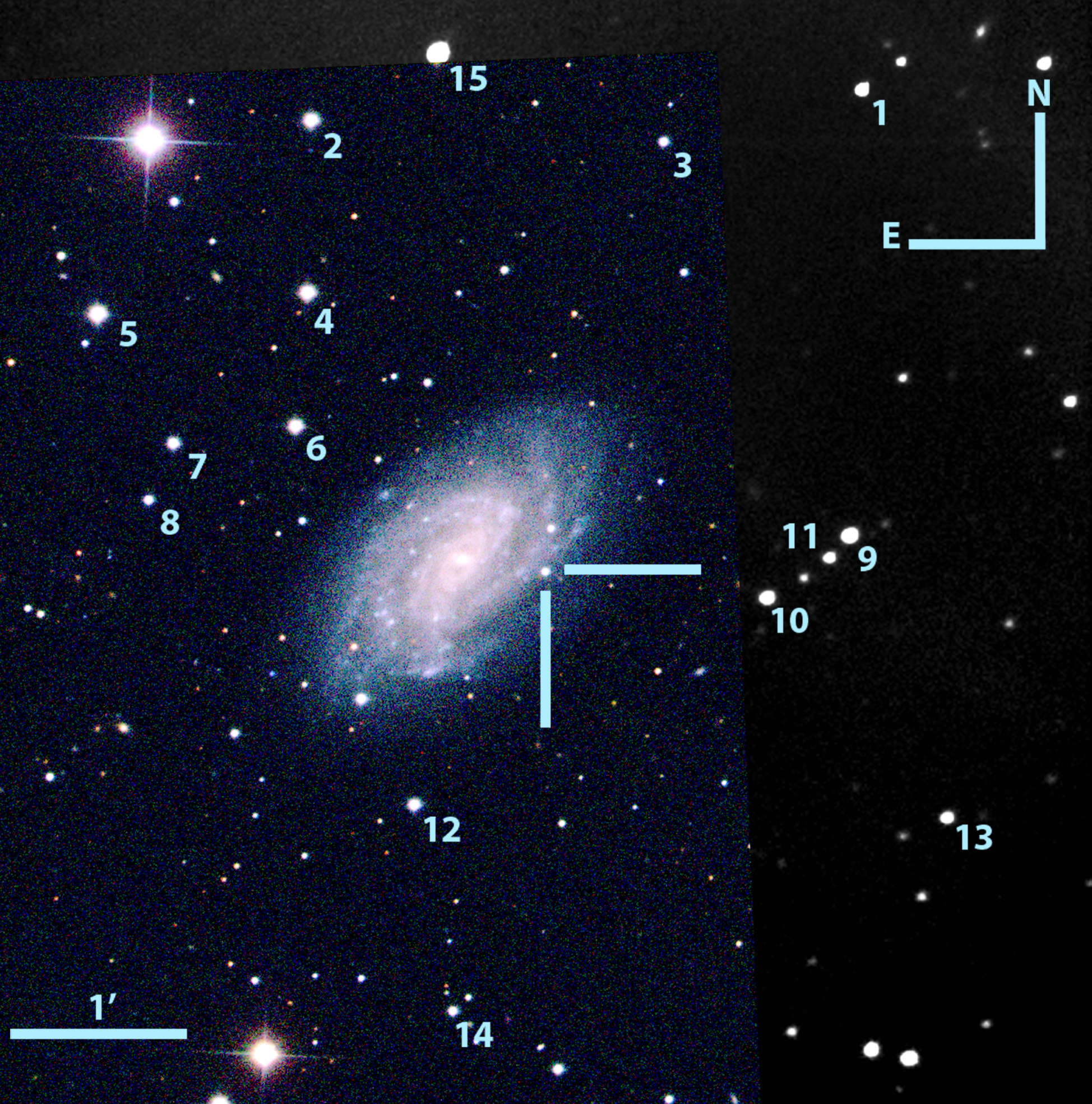}
\caption{\label{finder}Color composite of $g$-, $r$-, and $i$-band SDSS 
images of \cb\ overlain on a KAIT $V$-band image. The SDSS 
observation serendipitously caught the SN on 4 May 2000, shortly 
after discovery. The SN is located at a projected distance of 4.3\,kpc 
from the nucleus of its host galaxy, IC\,1158.}
\end{figure}

Photometric observations are given in Table \ref{phot_table}. Data
were obtained using KAIT, the robotic 0.76-m reflecting telescope at
Lick Observatory, between 29.41 April and 15.14 September 2000, both
in $BVI$ filters and as unfiltered observations that were made as part
of the routine SN search. Following automatic on-site flat-fielding
and bias subtraction, images were reduced using the KAIT pipeline
\citep[see][]{ganeshalingam10}. Galaxy subtraction was performed using
template images from several months after the SN had faded beyond
detection, and the DAOPHOT package in IRAF\footnote[1]{The Image
  Reduction and Analysis Facility: distributed by the National Optical
  Astronomy Observatory, which is operated by the Association of
  Universities for Research in Astronomy (AURA) under cooperative
  agreement with the National Science Foundation.} was used to
perform point-spread function (PSF) fitting photometry on the SN as
well as on comparison stars in the field. Figure \ref{finder} shows
the SN, its host, and the comparison stars, whose magnitudes are
listed in Table \ref{stand_table}. All magnitudes are on the
Vega-based system.

We correct our data for Galactic extinction of $E(B - V)_{\rm Gal} =
0.114$ mag from \citet{schlegel98}, using the extinction law of
\citet{cardelli89}, with a slope of $R_V=3.1$. The unfiltered
photometry of SNe with KAIT is usually dominated by $R$-band light
\citep[e.g.,][]{li03}. This appears to be a good approximation for
\cb; the unfiltered data lie between the $V$ and $I$ light curves
throughout most of the lifetime of the SN. We therefore correct the
unfiltered data for Galactic extinction as though they were acquired
with an $R$-band filter. The distance to \cb\ is 30\,$\pm$\,7\,Mpc
\citep{springob09}, and its measured redshift is $z = 0.0064$. We find
the explosion date to be 21.5 April 2000 (JD 2,451,656) $\pm$ 4.1~d
by extrapolating the unfiltered light curve using a cubic fit to the
first five points.

Our light curve of SN\,1999em was first published by
\citet{leonard02a}. We adopt a total extinction value $E(B-V)_{\rm
  total} = 0.1$ mag from spectral fitting \citep{baron00} and a
Cepheid distance of 11.7\,$\pm$\,1.0\,Mpc \citep{leonard03}, with a
redshift of $z = 0.0024$. We assume the explosion date of 23.5 October
1999 found by \citet{jones09} using EPM.

The light curve of SN\,1987A \citep{hamuy88} is corrected for
extinction using $E(B - V)_{\rm Gal} = 0.075$ mag \citep{schlegel98}
and $E(B - V)_{\rm host} = 0.1$ mag \citep{woosley87}. We adopt a
distance of $50 \pm 5.2$\,kpc to the Large Magellanic Cloud
\citep{storm04} and an explosion date of 23.316 February 1987
\citep{arnett89}.


Two early-time spectra of SN\,2000cb, around days 7 and 9 after
explosion, were obtained using the FAST spectrograph
\citep{fabricant98} on the 1.5\,m Tillinghast telescope at the
F. L. Whipple Observatory (slit width $3''$, resolution $\sim 7$ \AA).  Beginning about 40\,d
after explosion, seven additional spectra were taken with the Kast
spectrograph on the Shane 3\,m reflecting telescope at Lick
Observatory \citep{miller93}. The first five were obtained using only
the red side of the dual-arm Kast spectrograph, and the last two were
taken with both sides. With a slit width of $2''$, the typical
resolution (full width at half-maximum intensity) was $\sim 6$~\AA\ on
the blue side ($\lambda \lesssim 5500$~\AA) and $\sim 8$~\AA\ on the
red side. The slit was in all cases aligned close to the parallactic
angle \citep{filippenko82} in order to minimise the effects of
atmospheric dispersion. Table \ref{spec_table} provides a log of the
spectroscopic observations.

All spectra were reduced using techniques detailed by \citet{foley03}
and \citet{matheson08}, among others. IRAF CCD processing was
performed, and the data were sky subtracted and extracted optimally
\citep{horne86}. The observations were wavelength calibrated by
fitting low-order polynomials to arc-lamp spectra. Night-sky lines
were cross-correlated with a template sky spectrum and small
wavelength shifts were applied when appropriate. Using the spectra of
standard stars, the data were then flux calibrated and corrected for
telluric absorption lines \citep{wade88,bessell99,matheson00}. 
Information for the Kast data comes from our SN database 
(Silverman et al., in prep.).

\vspace{0.8cm}

\begin{figure}
\centering
\includegraphics[width=3.0in]{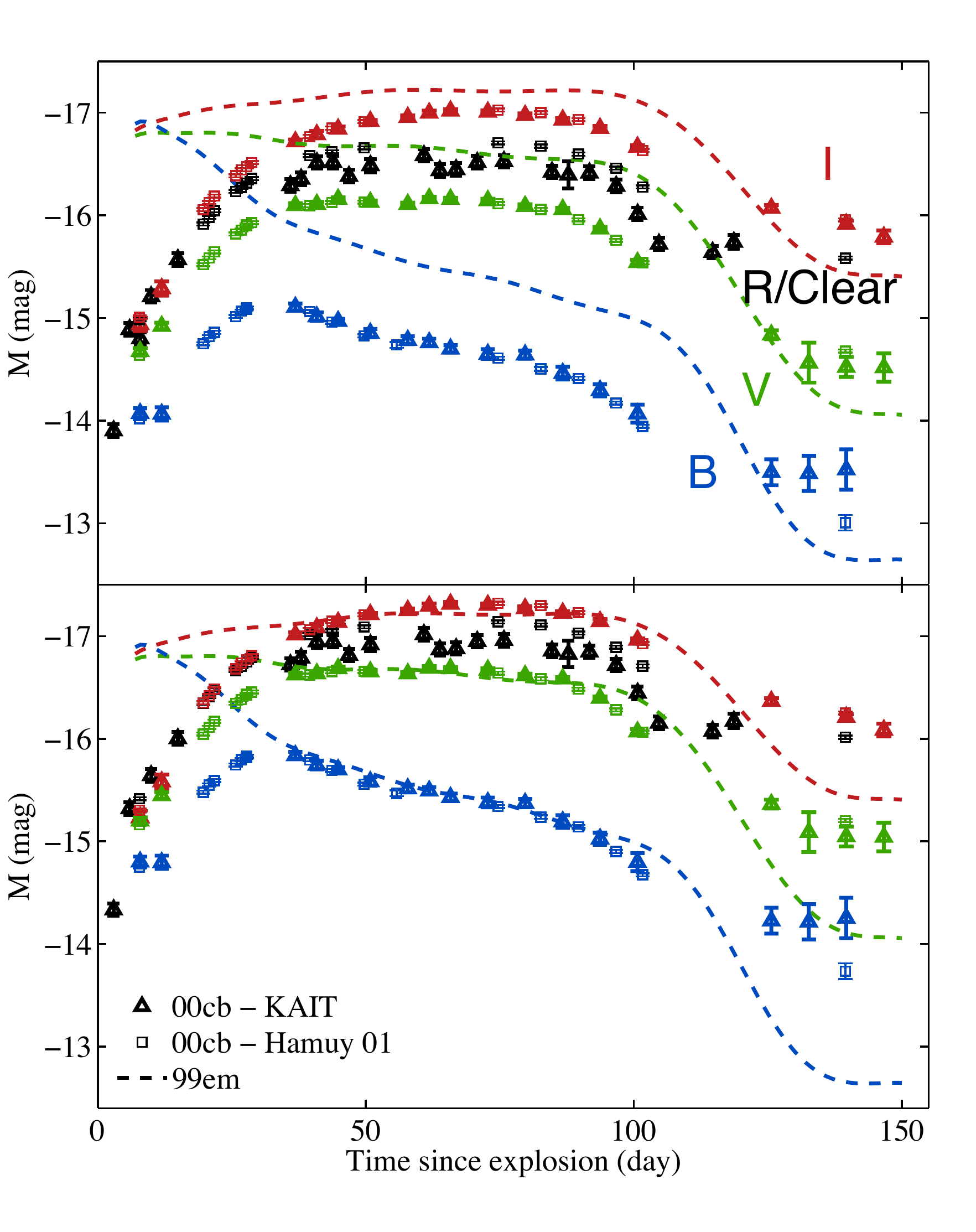}
\caption{\label{lc-99em} {\it Top:} Light curves of SNe\,2000cb and
  1999em in $B$, $V$, and $I$, assuming no host-galaxy extinction for
  \cb. {\it Bottom:} The same light curves, but with \cb\ matched to
  SN\,1999em using $E(B - V)_{\rm host} = 0.2$ mag and $R_V =
  2.6$. While assuming significant dust extinction can cause the light
  curves to match for a fraction of the SN lifetimes, it is
  inconsistent with our spectroscopic limits and the location of the
  SN within its host galaxy. Note that there is a systematic
  uncertainty in the scaling, mostly due to the distance to \cb, that
  is not known to better than $\sim 0.5$ mag.}
\end{figure}

\begin{figure}
\centering
\includegraphics[width=3.5in]{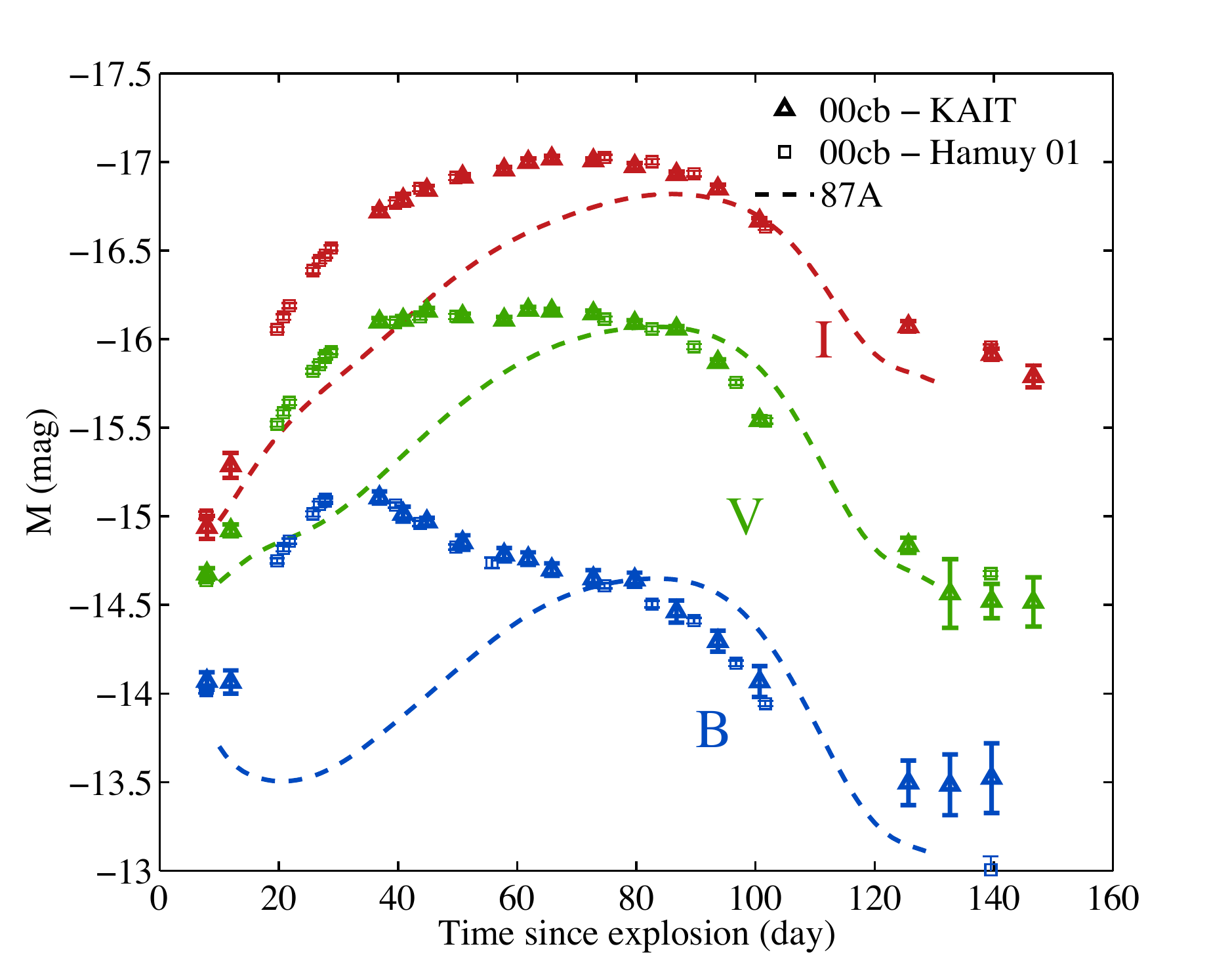}
\caption{\label{lc-87A}Similar to Figure \ref{lc-99em}, but comparing
  the light curves to those of SN\,1987A. While there is some qualitative
  resemblance in the photometric evolution, particularly at later
  times, the SNe clearly develop on different time scales. The
  uncertainty for SN\,1987A is dominated by 0.1 mag associated with
  the distance-modulus correction.}
\end{figure}

\begin{figure}
\centering
\includegraphics[width=3.0in]{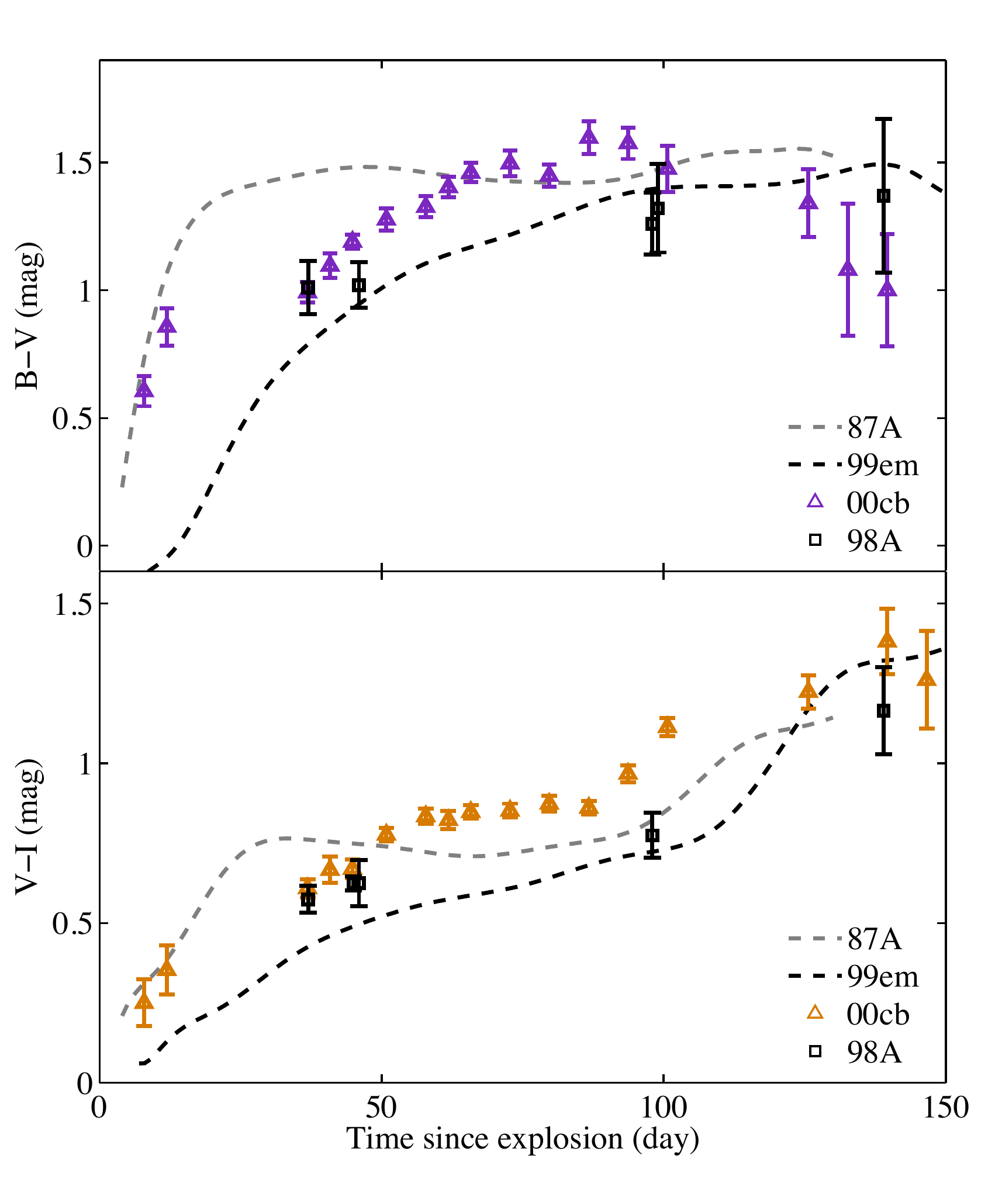}
\caption{\label{colors}Colour curves of \cb\ compared with those of
  SNe\,1999em, 1998A, and 1987A. The three objects exhibit similar
  basic trends but also some significant differences.}
\end{figure}

\section{Photometry}\label{s:phot}

In the top panel of Figure \ref{lc-99em}, we compare the light curves
of \cb\ and SN\,1999em. \cb\ is fainter by 2--3 mag in each band
at early times and then ascends during the first 30\,d into the
plateau. During the photospheric phase, \cb\ follows this prototypical
SN\,II-P reasonably well until day 90, when it begins to drop off the
plateau. This happens about 20\,d earlier than for SN\,1999em, but the
decline is not as rapid. As a result, \cb\ is overluminous by up to
$\sim 1$ mag, depending on the band, from day 110 until the latest
observation at day 147. Our photometric data are listed in Table
\ref{phot_table}.

\begin{figure*}
\centering
\includegraphics[width=7in]{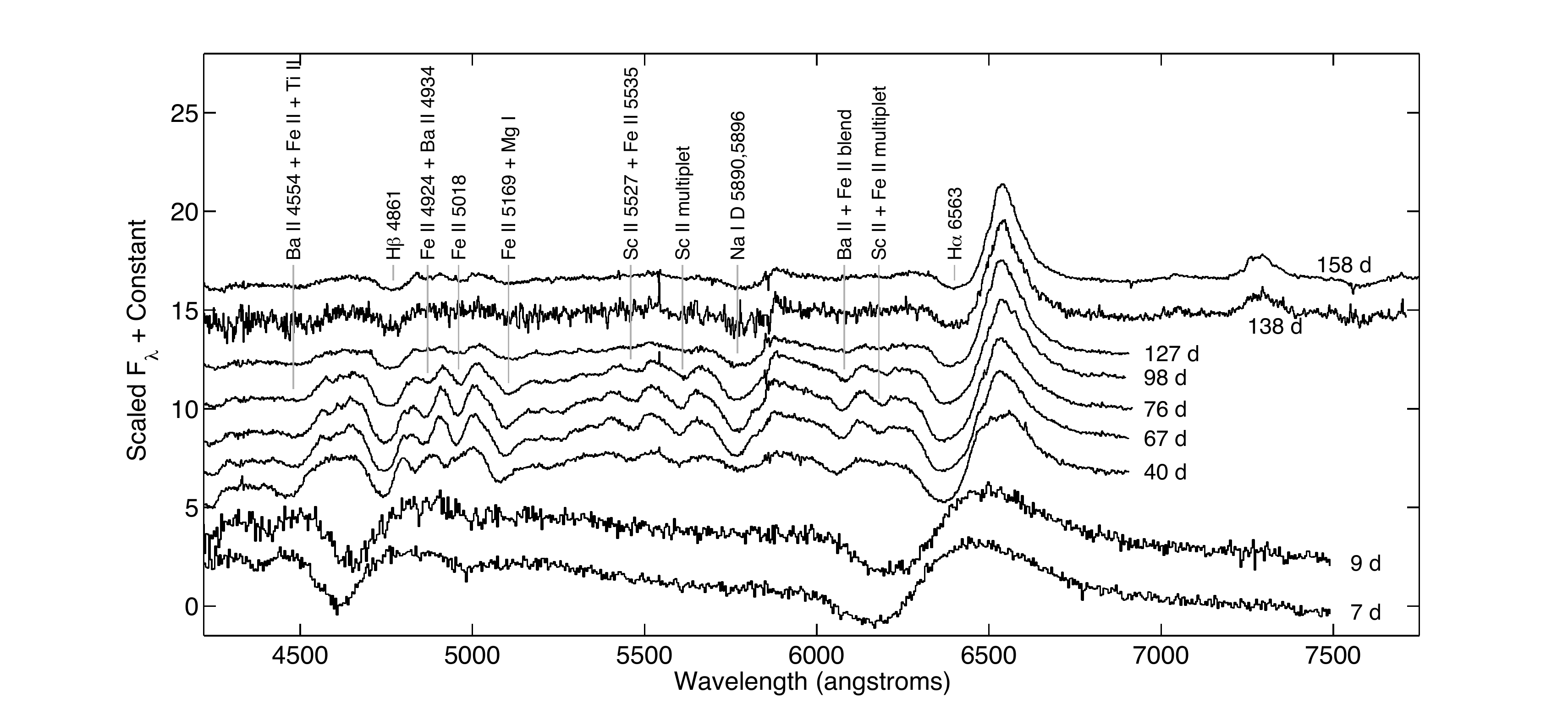}
\caption{\label{allspec}Spectroscopic evolution of \cb, with prominent
  features identified.}
\end{figure*}

The assumption that SNe\,II-P have essentially the same colour
evolution during the photospheric phase is one method often used to
determine reddening \citep[e.g.,][]{leonard02b}. Such a fit requires a
value of $E(B - V)_{\rm host} = 0.2$ mag with $R_V = 2.6$ to be
applied to \cb, as shown in the bottom panel of Figure
\ref{lc-99em}. This allows us to match the light and colour curves of
the two objects reasonably well during the plateau phase, roughly
between days 30 and 90 after explosion. However, such a significant
extinction value is inconsistent with the absence of detectable
narrow Na\,I\,D absorption (see \S\ref{s:spec}), and it seems unlikely
considering the position of the SN near the outskirts of its host
galaxy. We therefore assume throughout the rest of this paper that
\cb\ suffered no host-galaxy extinction, though the presence of some 
extinction would not appreciably alter any of our conclusions.

In Figure \ref{lc-87A}, we compare the photometric evolution of
\cb\ and SN\,1987A. While the SNe show similar qualitative trends such
as a slow rise time, a plateau-like phase, and a shallow dropoff (as
also noted by \citealt{hamuy01b}), they clearly behave differently.

Figure \ref{colors} shows $B - V$ and $V - I$ colour curves of
\cb\ compared with those of SNe\,1999em, 1998A, and 1987A. At early
times, \cb\ appears to follow SN\,1987A, and both are redder in $B -
V$ than SN\,1999em beyond anything that could be ascribed to
extinction, which indicates a lower effective temperature. In $B - V$,
\cb\ joins the bluer objects from about day 30 until day 60, when the
colours of all four are comparable. Until day 100 or so, \cb\ is
slightly redder than SNe 1999em and 1998A, but perhaps not
significantly. As it falls off the plateau, \cb\ begins to evolve
quickly toward the blue. In $V - I$, \cb\ again tends to be redder
than the three comparison SNe between days 50 and 100, but overall
these objects all show broadly similar colour evolution.

\section{Spectroscopy}\label{s:spec}

\begin{figure}
\centering
\includegraphics[width=3.0in]{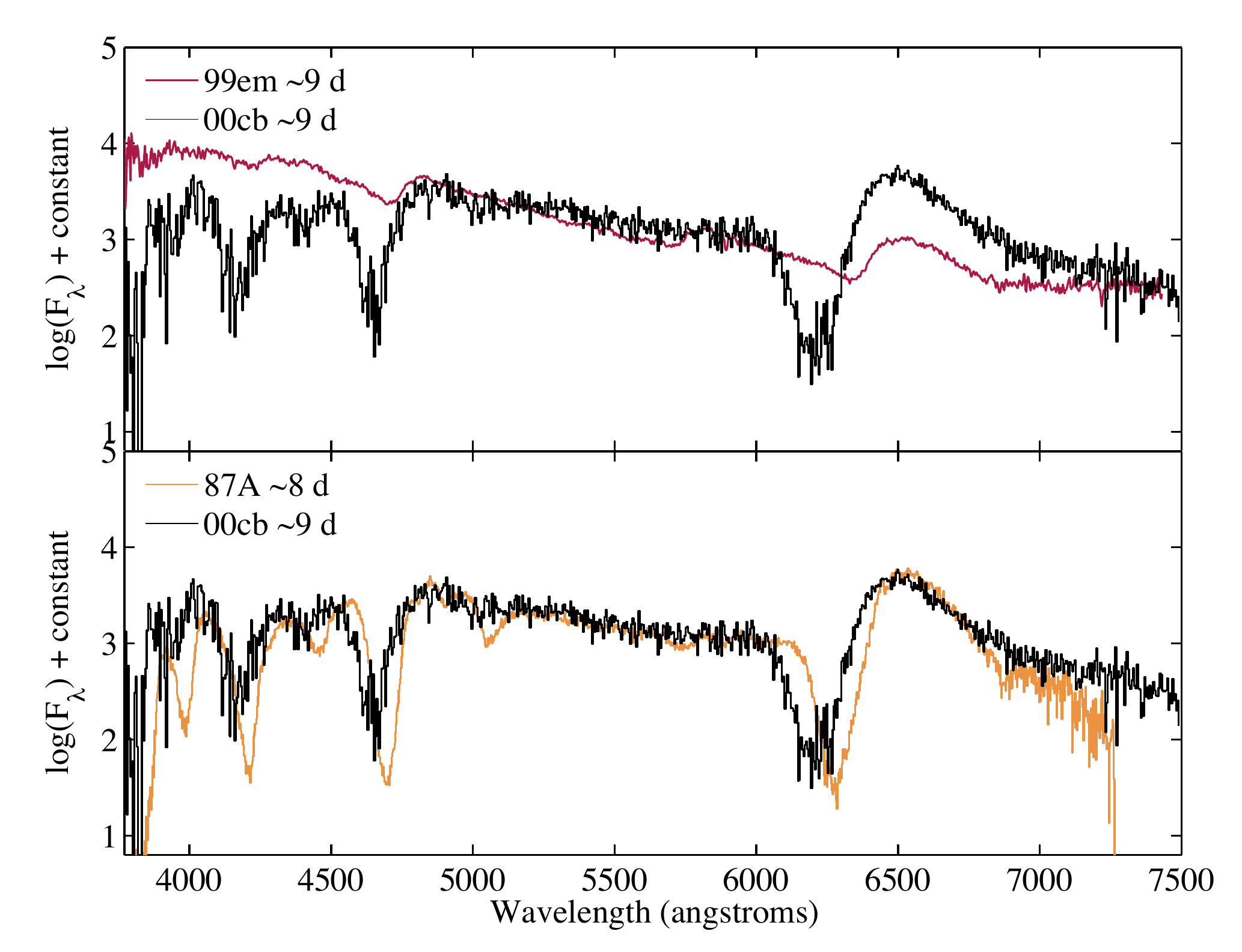}
\includegraphics[width=3.0in]{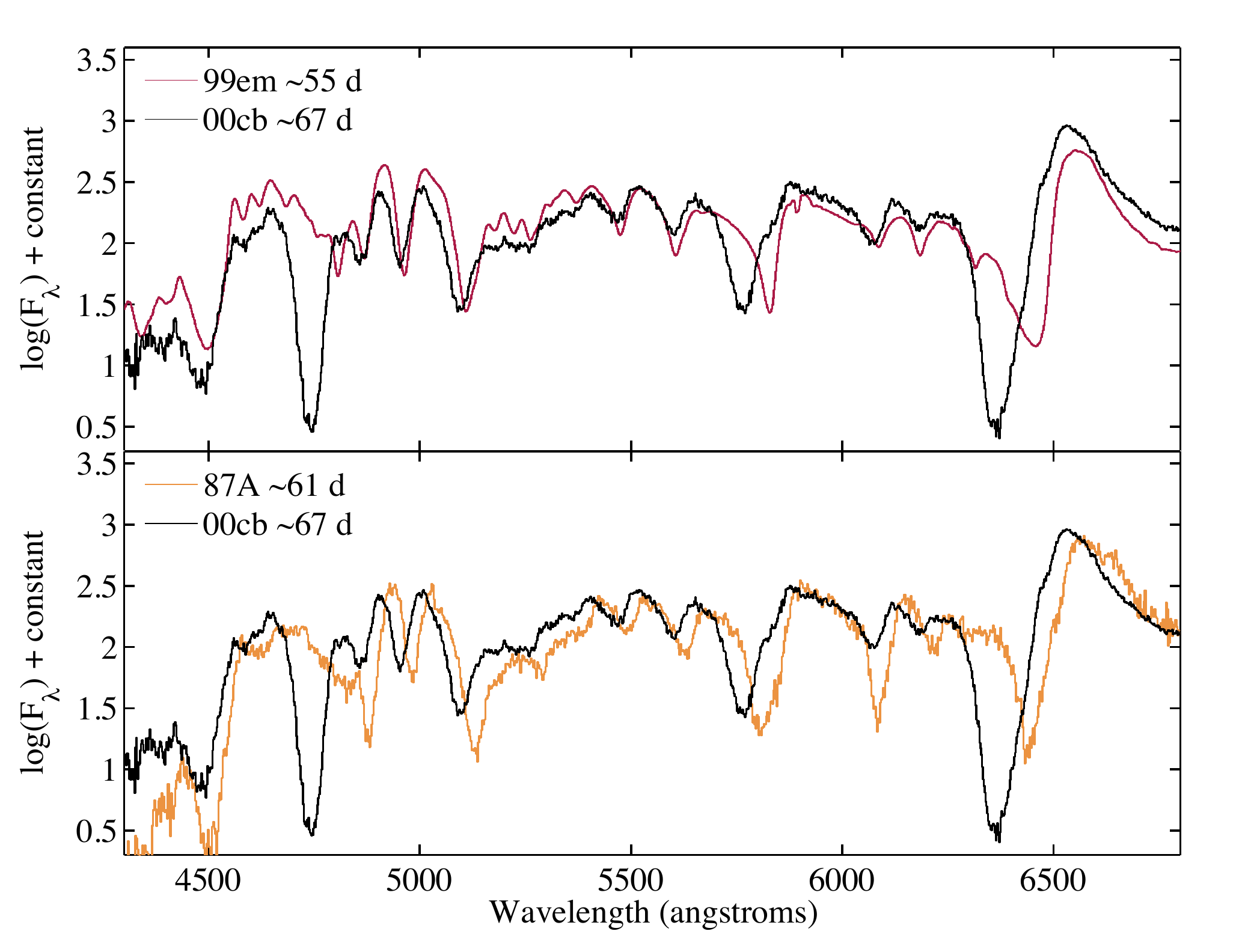}
\includegraphics[width=3.0in]{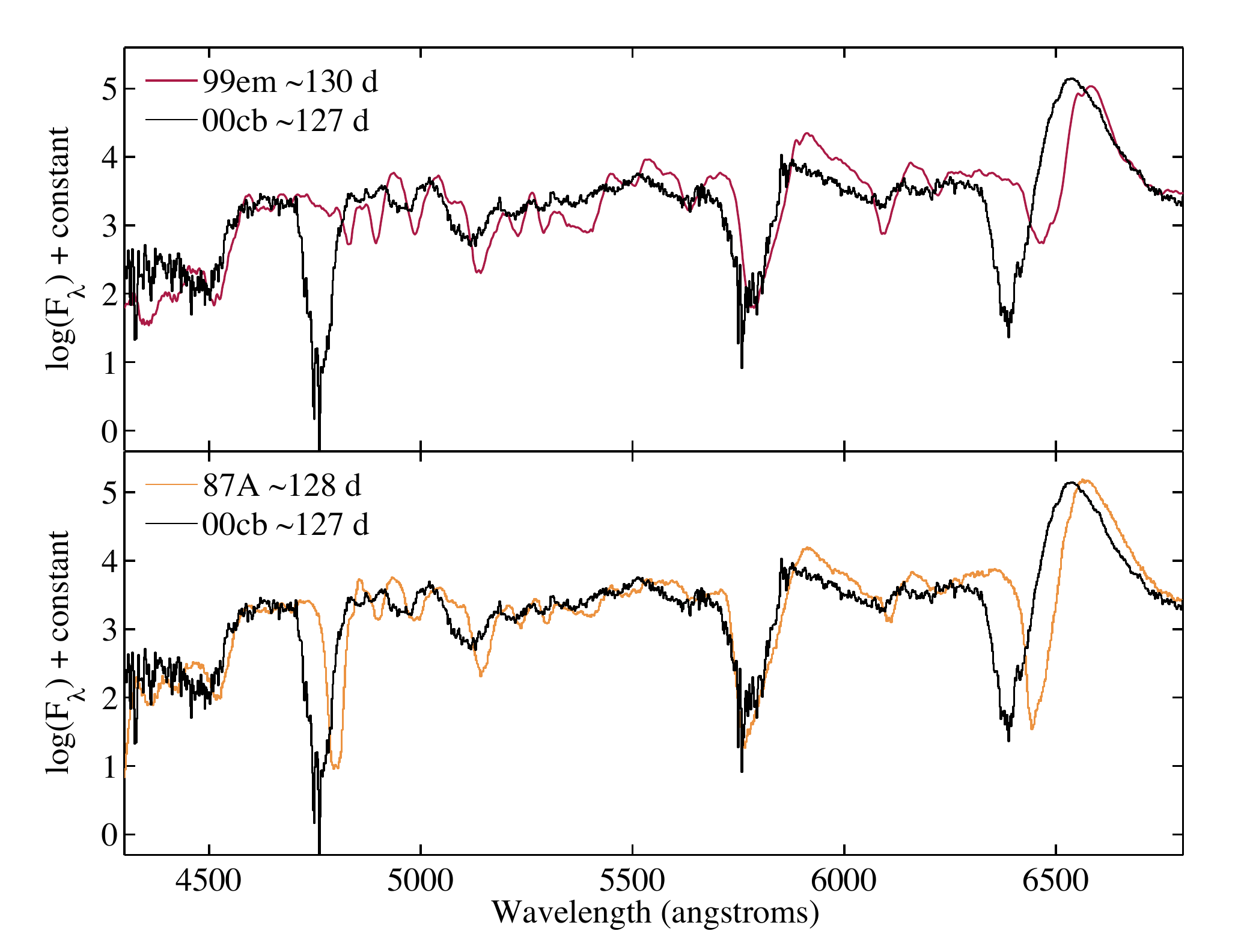}
\caption{\label{compspec} Optical spectra of \cb\ at days 9, 66, and
  126 compared with spectra of SN\,1999em and SN\,1987A at similar
  phases. Each is shown in the supernova's rest frame. Note the high
  expansion velocities of \cb, the early-time similarity of \cb\ to SN\,1987A,
  and the prominence of the H$\alpha$ absorption feature in the day 67 spectrum.}
\end{figure}

Spectra obtained between days 7 and 157 after explosion are shown in
Figure \ref{allspec}. All exhibit features characteristic of SNe\,II
during the photospheric phase, including strong H$\alpha$ P-Cygni
profiles as well as H$\beta$ absorption, Na\,I\,D
$\lambda\lambda$5890, 5896, Fe\,II $\lambda\lambda$4924, 5018, 5169
(hereafter, Fe triplet), and perhaps Ba and Sc lines. A number of
these absorption features are labeled in the figure based on line
identifications for SN\,1999em \citep{leonard02a} and SN\,1998A
\citep{pastorello05}.

\begin{figure}
\centering
\includegraphics[width=3.5in]{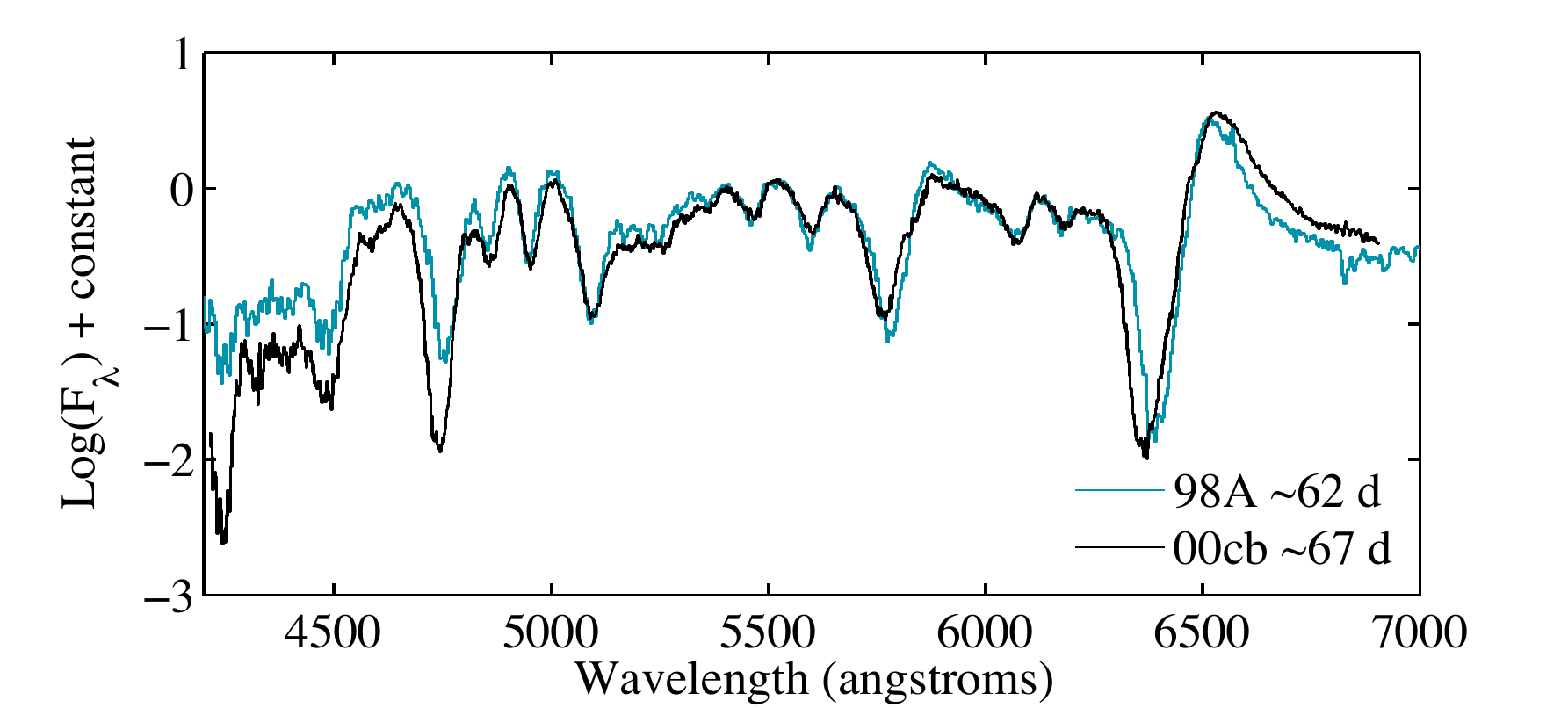}
\caption{\label{s98a} Comparison of the SN\,1998A spectrum at $\sim
  62$\,d after explosion with that of \cb\ at a similar phase.}
\end{figure}

There are indications that the equivalent width of narrow Na\,I\,D
$\lambda\lambda$5890, 5896 absorption correlates with the amount of
dust extinction suffered by a SN, despite significant scatter (e.g.,
\citealt{barbon90,richmond94,turatto03}; see also Poznanski et al., in
prep.). Regardless of the scatter, there seems to be a correlation
between high extinction and Na\,I\,D absorption. We detect no
absorption in any of the individual spectra of \cb\ nor in a stack of
all nine spectra.

For the stack, we normalise the spectra, crop a region of
60~\AA\ around 5893~\AA, subtract a fitted line to remove any
continuum contribution, and combine the spectra. The resulting average
spectrum has a standard deviation of 5\% in flux, and yet we do not
detect any narrow Na\,I\,D absorption. For a typical resolution
element of about 5~\AA, we thus get a 3$\sigma$ limit of
0.47~\AA\ on the equivalent width, following Equation 4 of
\citet{leonard01}; this corresponds to $E(B-V) < 0.12$ mag
(using the relation of \citealt{barbon90}), or $A_V< 0.31$ mag
for the best-fitting value to the photometry of $R_V=2.6$. Such
a low $E(B-V)$ is inconsistent with the value 0.2 derived by
matching the photometry to SN\,1999em, and it indicates that \cb\ 
most likely suffered no significant host-galaxy extinction.

The spectra of \cb\ are overall very similar to those of typical
SNe\,II, but direct comparisons to SN\,1999em and SN\,1987A reveal
several significant differences. Figure \ref{compspec} shows
individual spectra of \cb\ plotted against spectra at similar phases
for each of these objects.

%

\begin{figure}
\centering
\includegraphics[width=3.0in]{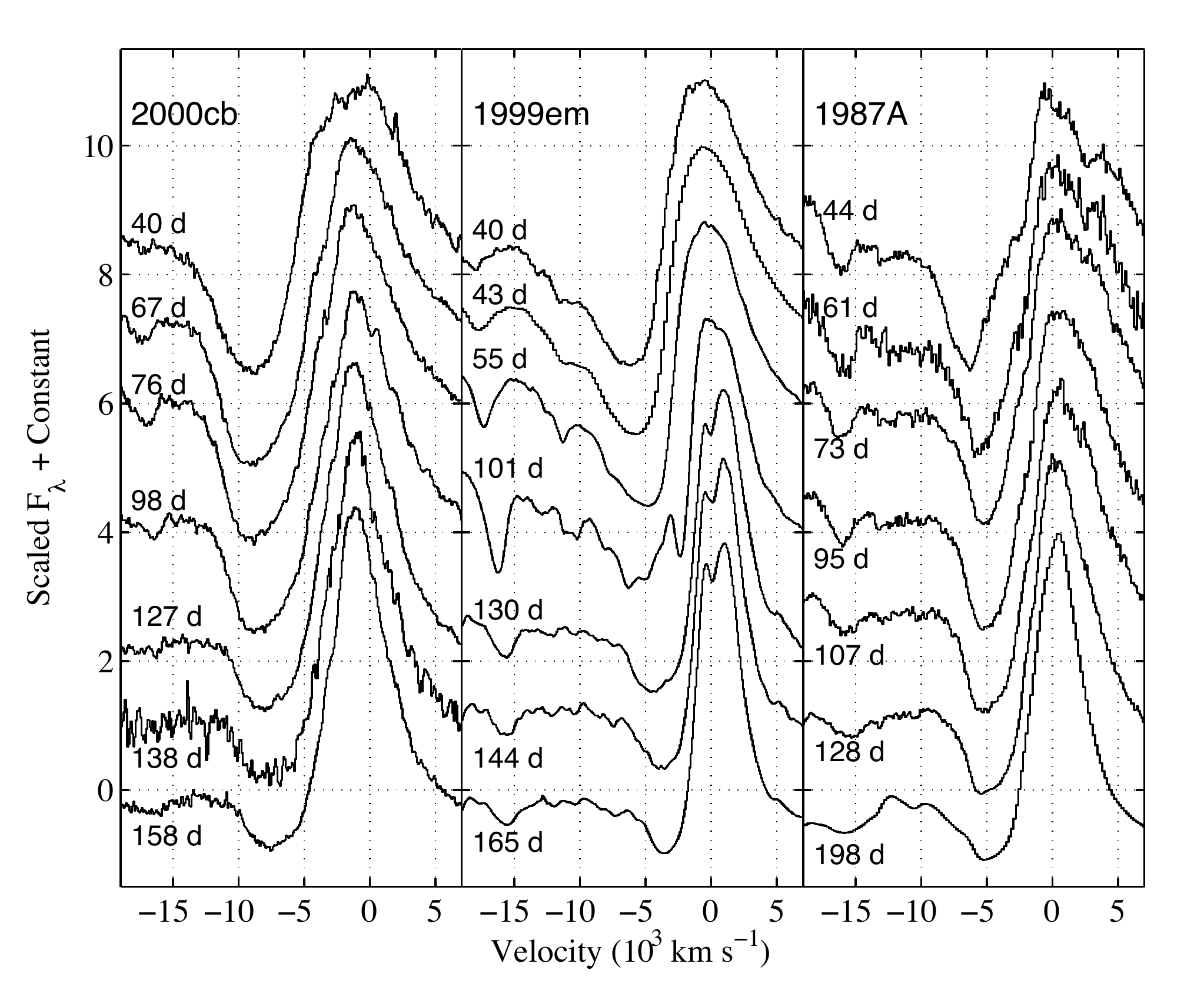}
\caption{\label{halpha}H$\alpha$ feature in spectra of SNe\,2000cb,
  1999em, and 1987A, showing the velocity evolution and line
  profiles. All spectra are scaled by their H$\alpha$ emission
  strength and corrected for redshift.}
\end{figure}

At early times, \cb\ closely resembles SN\,1987A: their continua match
well, and the strengths of the prominent hydrogen features are
comparable. Both are redder than SN\,1999em, which also has much
weaker features.  The features of \cb\ tend to be significantly
blueshifted compared to either object throughout their evolution in
our sample. Around day 67, all three objects are very similar
spectroscopically, although \cb\ has much deeper absorption troughs
for H$\alpha$ and even more strikingly for H$\beta$, which is barely
discernible in the other two objects. At about day 127, \cb\ again
matches both comparisons well; however, most of the Fe (and other
metal) absorption lines are weaker. \cb\ still exhibits strong
hydrogen compared to SN\,1999em, but at this phase SN\,1987A shows
features of comparable depth. The equivalent widths of the hydrogen
absorption lines for \cb\ and SN\,1987A exceed those of SN\,1999em by
a factor of 8--10 at early times, and at later times measurements of
\cb\ are about twice as great as those of SN\,1987A, while features of
SN\,1999em are even weaker.  We also note that the feature at about
5600~\AA\ in \cb, identified as Na\,I\,D $\lambda\lambda$5890, 5896,
has a profile that differs from that of both SNe\,1999em and 1987A in
the day 67 comparison, but matches reasonably well at later times. This
feature may be a blend with He\,I $\lambda$5876.

SN\,1998A, identified by \citet{pastorello05} as a higher-energy
analog of SN\,1987A, is a very close spectroscopic match to \cb, as
shown in Figure \ref{s98a}; its spectra are almost identical at all
other available phases as well. The depths of the hydrogen features
are comparable, while SNe\,1999em and 1987A differ greatly. It is interesting
to note the spectroscopic similarity between SN\,2000cb and SN\,1998A
in spite of the fact that they are not nearly as close photometrically,
the light curve of SN\,1998A being nearly identical to that
of SN\,1987A but brighter by about 0.7 mag.

Figure \ref{halpha} illustrates the evolution of the H$\alpha$ profile
beginning at $\sim 40$\,d after explosion.  In addition to the higher
velocity and velocity dispersion of the absorption minimum, the
emission peak of \cb\ is blueshifted significantly more than those of
the other two SNe. The H$\alpha$ emission profile also exhibits a
slight excess on the blue side. This structure in the P-Cygni profile
is reminiscent of the ``Bochum event'' observed in SN\,1987A
\citep{hanuschik87,utrobin95}. A similar feature, tentatively attributed to the
same phenomenon by \citet{leonard02a}, also appears briefly in the
profile of SN\,1999em, but it is not present over as much time in that
object's evolution (only at $\sim 101$\,d in our spectra).

\begin{figure}
\centering
\includegraphics[width=3.5in]{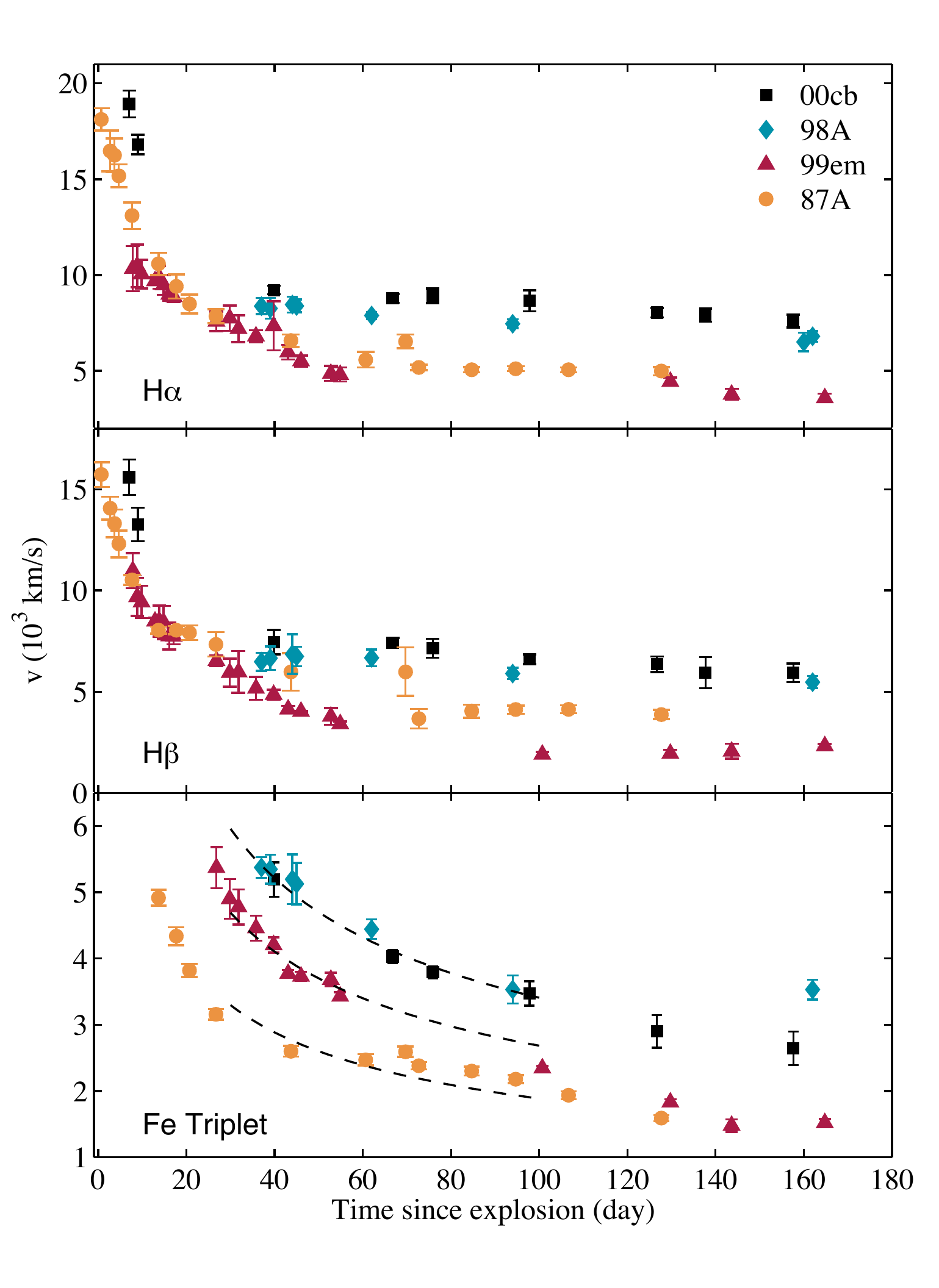}
\caption{\label{velocities}Velocity evolution of \cb\ compared to SNe
  1999em, 1998A, and 1987A as measured from H$\alpha$ (top), H$\beta$
  (middle), and averaged Fe triplet (bottom). The dashed lines follow
  the relation found by \citet{nugent06}.}
\end{figure}

We have measured the velocities of the most prominent absorption
features of all four SNe, as shown in Figure \ref{velocities}. The Fe
triplet velocities (shown in the plot as the average of the velocities
measured for Fe\,II $\lambda\lambda$4924, 5018, 5169) of \cb\ and
SN\,1998A are similar and typically exceed those of SN\,1999em by
$\sim 1000$--1500 km\,s$^{-1}$ and those of SN\,1987A by $\sim
2000$--2500 km\,s$^{-1}$. However, they appear to follow roughly the
empirical relation (indicated for each object by dashed lines in the
same figure) found by \citet{nugent06} for SNe\,II-P: $V(50) =
V(t)(t/50)^{(0.464 \pm 0.017)}$, where $V(50)$ is the velocity 50\,d
after explosion and $V(t)$ is the velocity at some time $t$ in
days. The H$\alpha$ and H$\beta$ absorption velocities of \cb\ are
more remarkable. H$\alpha$ absorption velocities exceed 18,000
km\,s$^{-1}$ within 10\,d after explosion and remain $\sim 3000$--5000
km\,s$^{-1}$ greater than for both SNe 1999em and 1987A. The hydrogen
velocities measured for SN\,1998A are almost as high, lower than those
of \cb\ by about 500 to 1000 km\,s$^{-1}$. \citet{dessart08b} suggest
that in SN\,1987A-like objects, H$\alpha$ velocities overestimate
those of the photosphere by a factor of two at most phases due to non-
LTE effects. SN\,1987K \citep{filippenko88} and SN 2003bg
\citep{hamuy09} also had very high H$\alpha$ absorption velocities at
early times. However, the spectra of these objects soon developed He~I
lines, leading to the ``SN\,IIb'' designation. Our spectral sequence
in Figure 5 shows that \cb\ is not a member of this class.

\begin{figure}
\centering
\includegraphics[width=3.5in]{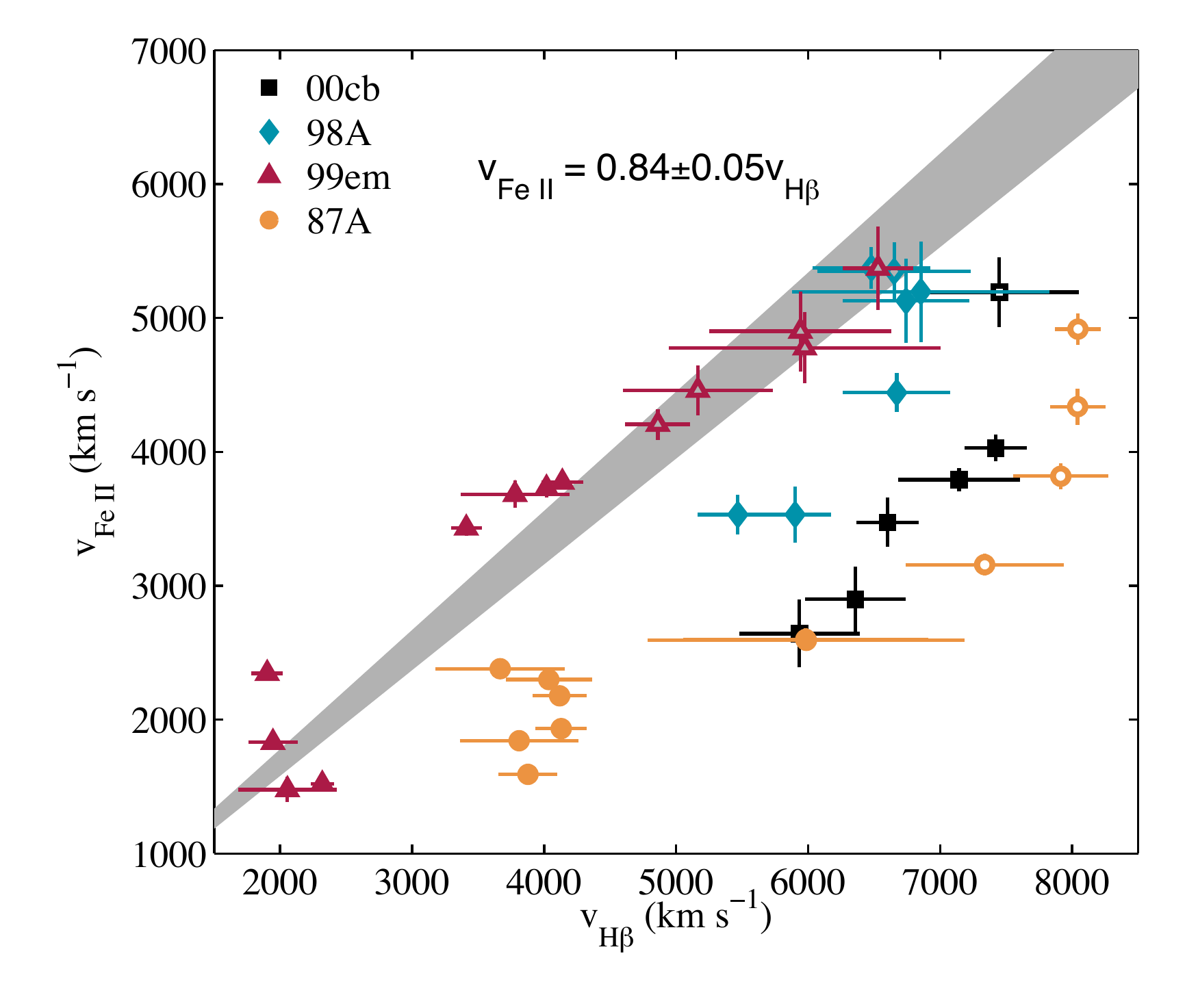}
\caption{\label{hbfe} H$\beta$ vs. Fe velocities for SNe\,2000cb,
  1998A, 1999em, and 1987A. The shaded region marks the 1$\sigma$
  relation found by \citet{poznanski10} for early-time
  spectra. Velocities measured earlier than day 40 (where the relation
  was defined) are marked with open symbols. Both SNe\,1987A and
  2000cb consistently lie well outside this region, while SN\,1999em
  remains near it even at later times. SN\,1998A is consistent with
  the relation at early times but subsequently diverges from it.}
\end{figure}

\citet{poznanski10} present a correlation between H$\beta$ and Fe\,II
velocities for H-dominated spectra of SNe\,II-P: $V_{\rm Fe} = (0.84 \pm
0.05)V_{{\rm H}\beta}$. Figure \ref{hbfe} shows this relation with our
velocity measurements; it was derived for spectra taken between 5 and
40\,d after explosion because of a tendency for H$\beta$ absorption to
become weak and blend with metal lines at later times. Our sample of
\cb\ contains only one usable spectrum within this timeframe (from day
40), but it is informative to examine the ratio of Fe and H velocities
at a wider range of epochs in the context of this correlation. 
SN\,1999em unsurprisingly follows this relation, as it was one of 
the SNe used to constrain it, especially at earlier times (and
therefore at higher velocities). \cb\ and SN\,1987A are clear outliers, 
while SN\,1998A deviates in a similar way but to a lesser extent.

\section{Bolometric properties and model fit}

\begin{figure}
\centering
\includegraphics[width=3.0in]{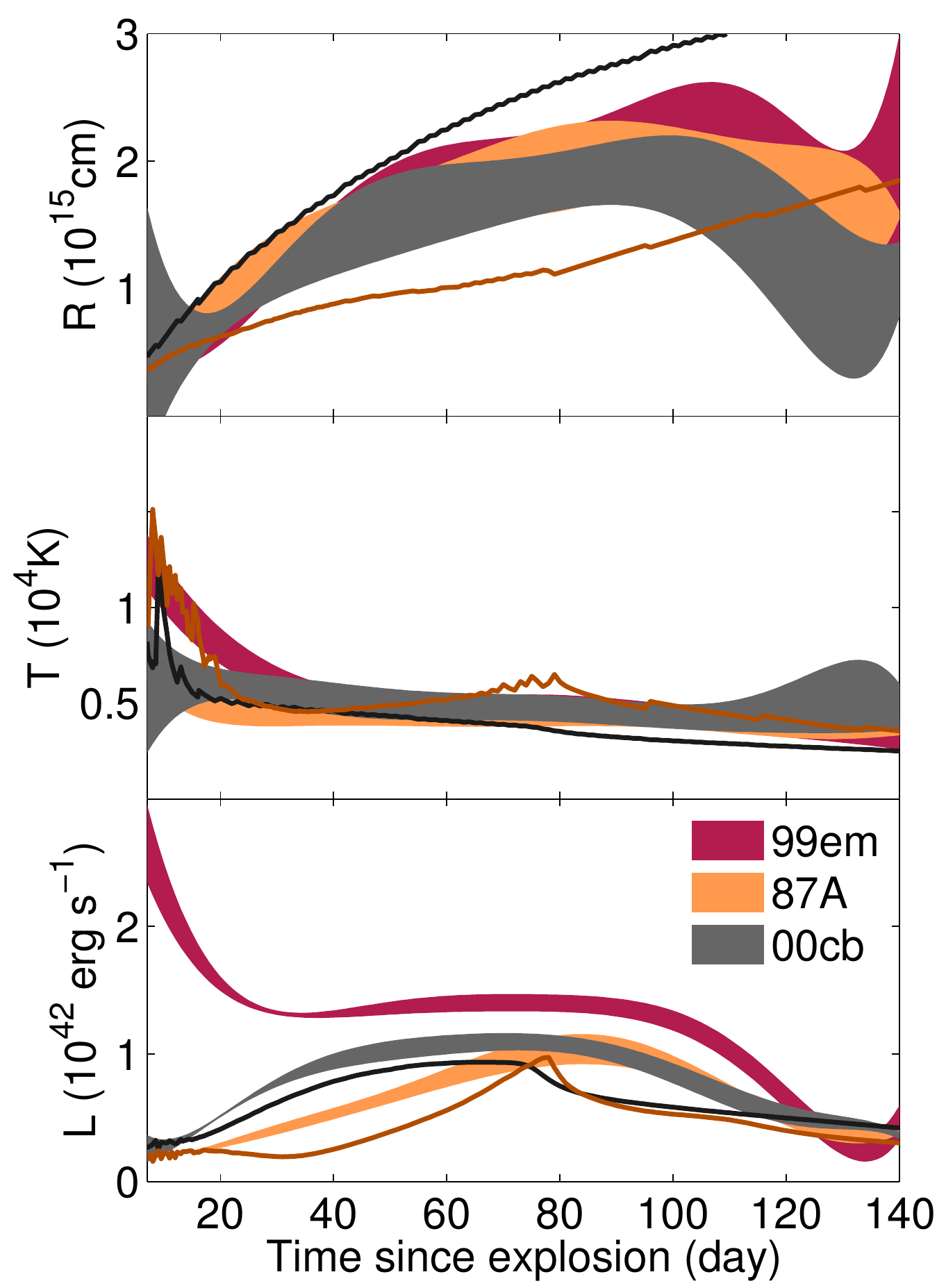}
\caption{Bolometric properties for SNe\,1987A, 1999em, and 2000cb,
  based on blackbody fits to the optical photometry. The thickness of
  the stripes marks 1$\sigma$ uncertainties. Lines are (approximate)
  model fits for SN\,2000cb and SN\,1987A generated using the code
  presented by \citet{young04} using $E
= 1$\,B, $R = 3 \times 10^{12}$\,cm, and $M_{\Nifs} =
0.07\,\Msun$ for SN\,1987A; and $E = 2.0$\,B,
$R = 3 \times 10^{12}$\,cm, and $M_{\Nifs} =
0.1\,\Msun$ for SN\,2000cb.\label{f:bol}}
\end{figure}

The first question posed by SN\,2000cb is whether its light curve was
powered primarily by the energy deposited in the initial explosion (as
in typical SNe\,II-P) or by the subsequent radioactive decay of
\Nifs\ (as in SN\,1987A and stripped-envelope Type Ib/c SNe). The slow
rise time strongly disfavours the former possibility. The explosion
shock wave deposits thermal energy throughout the stellar envelope,
including the outermost layers, and thus radiation begins escaping
from the ejecta immediately after shock breakout. The peak in the
$B$-band light curve occurs not long thereafter, when the outer layers
have cooled to temperatures $\la 10,000$~K. Model light-curve
calculations of SNe\,II-P find rise times of $\sim 10$~d in the $B$
band \citep{eastman94,baklanov05,kasen09,dessart10a}, significantly shorter than
the $\sim 35$\,d $B$-band rise time of SN\,2000cb. It is possible that
varying the structure of the hydrogen envelope could lengthen the 
rise time, but detailed modeling would be needed to explore such
effects. It is far more likely that the explosion energy was mostly
spent on expansion from a small progenitor radius, analogously to
SN\,1987A.

In order to derive bolometric properties, we fit a blackbody spectrum
to the optical photometry of \cb, SN\,1987A, and SN\,1999em, with the
luminosity, radius, and temperature being the free parameters. The
resulting fits are shown in Figure \ref{f:bol}. We assume that the
three objects differ from perfect blackbodies to a similar
extent. This may introduce a systematic uncertainty in our absolute
values, but it allows for comparisons to be made between the SNe. We find
for \cb\ a bolometric rise time of about 70\,d and peak luminosity of
$10^{42}$\,erg\,s$^{-1}$. Our derived values for SN\,1987A and
SN\,1999em are generally consistent with previous estimates. While all
three SNe have similar radii at most times, SN\,1999em is clearly
hotter and brighter at early times (as expected from its colours),
similar to other well-studied SNe\,II-P (e.g., \citealt{dessart08}).
\cb\ and SN\,1987A reach a similar peak luminosity, but \cb\ reaches
it about 20\,d earlier.

These long rise times are readily explained by a radioactive energy
source, as \Nifs\ is typically concentrated toward the centre of the
ejecta and radiation needs time to diffuse to the surface. Using the
scaling relation for the effective diffusion time $t \propto (\kappa
M_{\rm env}/v)^{1/2}$ \citep{arnett79}, we can estimate the ejected
mass, $M_{\rm env}$, and the explosion energy of SN\,2000cb from those
inferred for SN\,1987A: $M_{\rm 87A} \approx 14~\Msun$, $E_{\rm 87A}
\approx 1.1~{\rm B}$ (\citealt{blinnikov00}; see also
\citealt{woosley88,shigeyama90,utrobin95}), where $1 \mathrm{B} =
10^{51}$ erg. The rise time of SN\,2000cb is only 20\% faster than
that of SN\,1987A, and the expansion velocities are a factor of $\sim
1.7$ higher. Adopting a similar opacity $\kappa$ for both SNe would
give an ejected mass of $M_{\rm env} \approx 16.5~{\rm M}_\odot$, and
an explosion energy $E \approx M_{\rm env}v^2 \approx
4$~B. Preliminary results from PHOENIX modeling indicate that the
spectra of \cb\ are compatible with an ejecta mass of $M_{\rm env}
\approx 10~{\rm M}_\odot$ and the same explosion energy of 4\,B (Baron
et al., in prep.). These values are significantly higher than the
hydrogen envelope masses and energies typically inferred for SNe\,II-P
($M_{\rm env} \approx 8~\Msun$, $E \approx 1$\,B), though
\citet{utrobin07} do find higher envelope masses ($\sim 18~{\rm
  M}_\odot$) for both SN\,1987A and SN\,1999em.


Assuming a radioactive energy source, we can also estimate the mass of
\Nifs\ ejected in the explosion using Arnett's law \citep{arnett79},
which states that the luminosity at peak equals the instantaneous rate
of radioactive energy deposition. The luminosity we measure implies an
ejected \Nifs\ mass for SN\,2000cb of $\sim 0.14~\Msun$; however,
because of radiative diffusion this is actually an upper bound, as
some of this energy was deposited earlier in the light curve. If we
interpret the slope of the light curve at $t > 120$\,d as being due to
$^{56}$Co decay, then we find a \Nifs\ mass of $0.1 \pm
0.02~\Msun$. This matches the highest values measured from the decay
rates of SNe\,II-P at nebular phases \citep{hamuy03},
while the inferred value for SN\,1987A is $0.07~\Msun$.

The lack of a thermal (i.e., nonradioactive) contribution to the light
curve of SN\,2000cb --- or at the very least its dimness relative to
SN\,1999em at early times --- provides an upper limit on the
progenitor star radius. Both analytic and numerical models find that
the thermal luminosity of SNe\,II-P on the plateau scales roughly as
$L \propto E^{5/6} M^{-1/2} R^{2/3}$, where $R$ is the initial stellar
radius \citep{popov93,kasen09}. Comparing our above estimates of the
properties of SN\,2000cb to those inferred for SN\,1999em ($E_{\rm
  99em} \approx 0.7~{\rm B}$, $M_{\rm 99em} \approx 15~\Msun$, $R_{\rm
  99em} \approx 1000 R_\odot$; \citealt{baklanov05}), we find that the
lower bolometric luminosity of SN\,2000cb and its significantly higher
energy imply an initial stellar radius 10--15 times smaller, $R
\approx 5 \times 10^{12}$~cm, similar to that of SN\,1987A, $R_{\rm
  87A} \approx 3 \times 10^{12}$~cm \citep{woosley88}. Using the
values for SN\,1999em as found by \citet{utrobin07} gives similar
results. These scalings have been determined under the assumption that
the photosphere is near the recombination temperature of hydrogen,
$T_{\rm rec} \approx 6000$~K, which is consistent with our best-fitting
blackbody temperature during the plateau.

We use the one-dimensional Lagrangian radiation-hydrodynamics code
presented by \citet{young04} in order to model the light curves of
\cb\ and SN\,1987A. The pre-SN model for both objects is taken from
\citet{woosley95}, with an initial mass of $20~\Msun$. The code
assumes spherical symmetry and that the neutron star has a mass set as
an inner gravitational boundary. The simulation is initiated by
artificially placing thermal and kinetic energy in the inner zones,
extracted from the formation of the neutron star. The energy injection
produces a density discontinuity that propagates as an outward-moving
shock wave that takes several hours to exit the star. The simulation
follows the progression of the SN material as it expands and releases
the energy deposited by the shock.

Figure \ref{f:bol} shows our fits to SN\,1987A and SN\,2000cb. For
both SNe we reproduce the time scales and approximate maximum in the
light curves.  Included in the simulation is a calculation of the
deposition of energy from the decay of radioactive $^{56}$Ni. One of
the model parameters used to fit the light curve is the mixing of the
radioactive material within the ejecta. The more extensive the Ni
mixing, the sooner the effects of the heating will be observed. For
SN\,2000cb, we find it necessary to mix the $^{56}$Ni uniformly out to
$15\,\Msun$ of the ejecta. This allows an early heating of the
material and subsequent earlier peak in the light curve. In addition,
since the code does not take into account nonthermal opacity sources,
an opacity floor is used to simulate continued opacity due to
nonthermal ionization. The effect on the opacity of the large mass of
$^{56}$Ni and its substantial mixing is emulated by a higher opacity
floor.  In our model for SN\,1987A, more substantial mixing would help
smooth the sharp peak in the luminosity.

Fitting the velocity of the Fe lines in SN\,2000cb requires an
explosion energy of almost twice that of SN\,1987A. As a consequence
of the fast expansion, the debris becomes thinner faster, so a larger
ejected mass of $17.5\,\Msun$ is necessary to harness the radioactive
heating. This is 10\% higher than our fit to SN\,1987A. The radii and
temperatures we obtain are qualitatively consistent with the blackbody
fits to the photometry except at early times, where the temperature
from the model is significantly too high --- similar to SN\,1999em but
unlike what we measure for these very red SNe. The basic input
parameters for SN\,1987A are $E_\mathrm{87A} = 1$\,B, $R_\mathrm{87A}
= 3 \times 10^{12}$\,cm, and $M_{\Nifs} = 0.07\,\Msun$. For \cb\ we find
$E_\mathrm{00cb} = 2.0$\,B, $R_\mathrm{00cb} = 3 \times 10^{12}$\,cm,
and $M_{\Nifs} = 0.1\,\Msun$. Here again we find that \cb\ had a
relatively small initial radius consistent with a BSG, a significant
explosion energy (although not as high as derived from the scaling
above), and a substantial amount of \Nifs. It seems plausible that the
relatively high \Nifs\ yield of SN\,2000cb is related to its higher
explosion energy, consistent with the correlation proposed by
\citet{nadyozhin03}.

While there is typically some degeneracy between the energy and
envelope mass derived for SNe (as we can often constrain only their
ratio), we find for \cb\ that both the mass and energy were
substantial. The massive ejecta are required in order to trap the
gamma rays emitted by the decaying \Nifs. A smaller envelope would
entail more escape, pushing our \Nifs\ mass estimates even higher.

A possibility which we have not taken into consideration is that
asymmetry in the explosion and ejecta may have played a significant
role for this object. This would require extensive modeling and is
beyond the scope of this paper; moreover, we would also need to
explain why most SNe\,II do not have such bizarre properties or any
sign of significant asphericity \citep[e.g.,][]{leonard01}.

\section{SN\,2005\MakeLowercase{ci} and the rate of BSG explosions}

\begin{figure}
\centering
\includegraphics[width=3.5in]{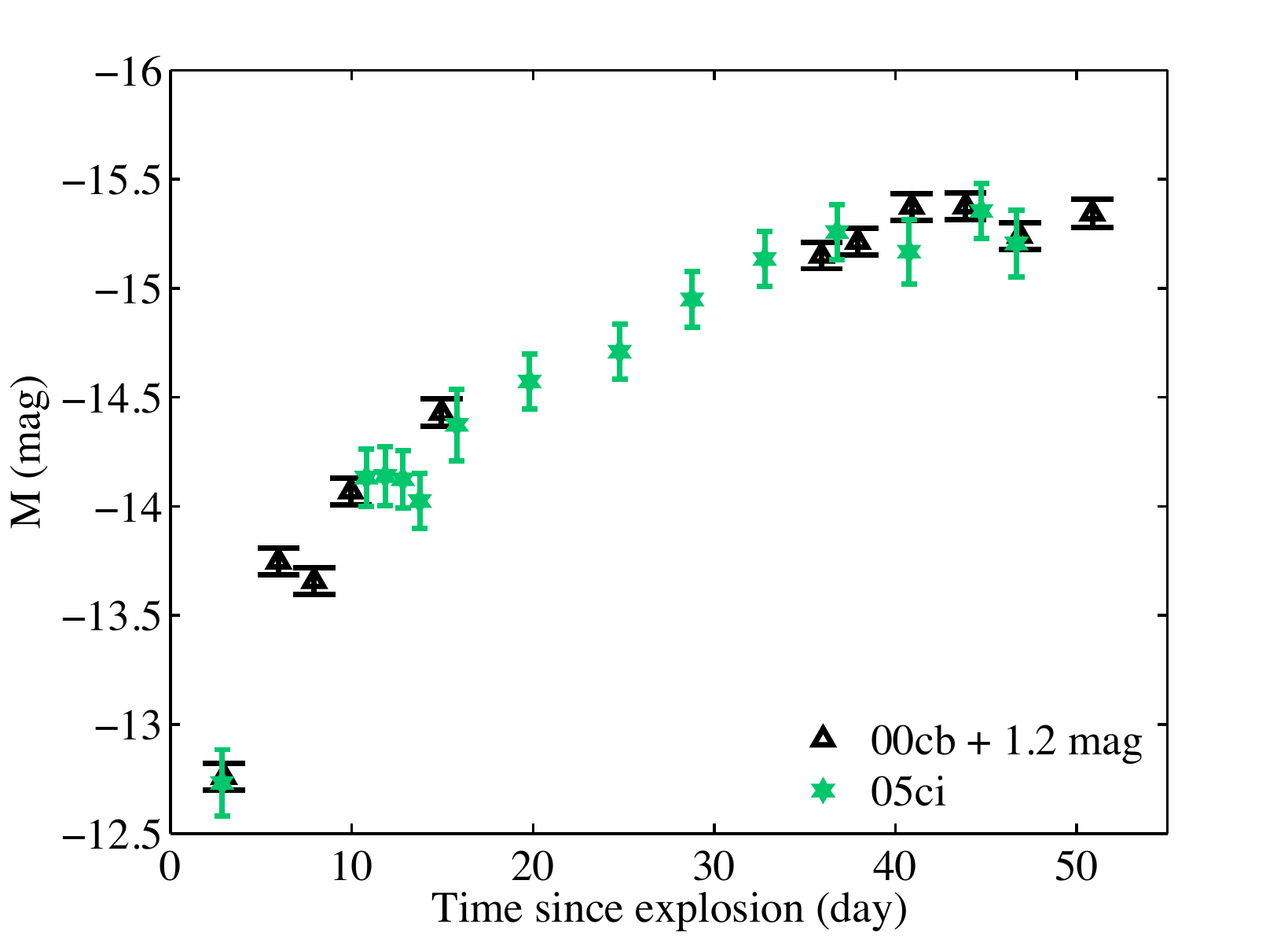}
\includegraphics[width=3.5in]{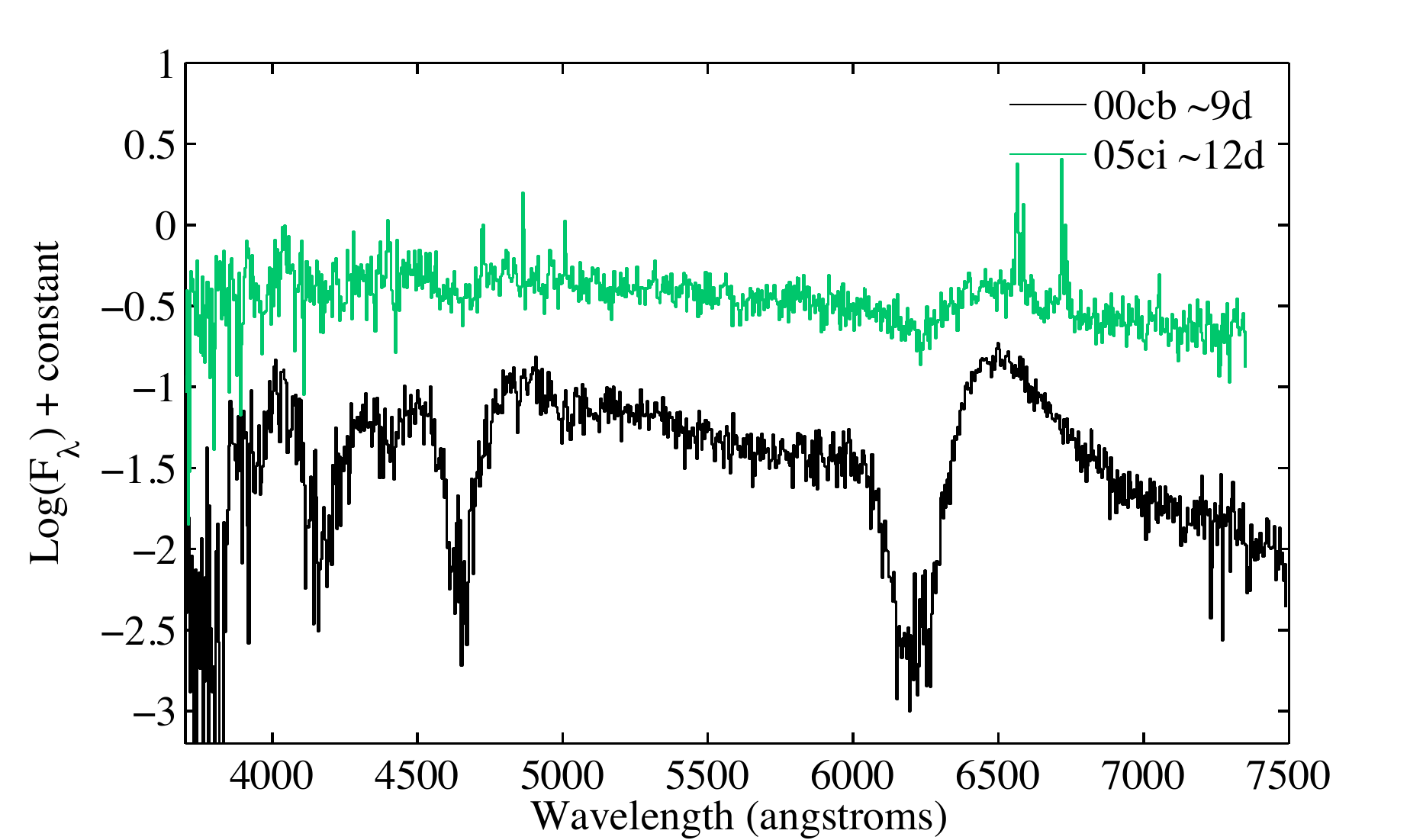}
\caption{Comparison of SNe\,2000cb and 2005ci. Unfiltered light curves
  (top) and spectra (bottom).\label{f:05ci}}
\end{figure}

While a few possible identifications have been made (e.g.,
\citealt{schmitz88,woodings98}; see a list in \citealt{pastorello05}),
the only published secure analog to SN\,1987A is SN\,1998A. Here, in
addition to SN\,2000cb, we report that SN\,2005ci
\citep{madison05,modjaz05} may be of the same ilk. The peculiarity of
this object was already noted by the Caltech Core-Collapse Project
\citep{arcavi09}. Unfortunately, we have only partial coverage of the
unfiltered light curve from KAIT (see Table
\ref{phot05ci_table}). Nevertheless, the resemblance to SN\,2000cb is
striking. While SN\,2005ci is fainter by $\sim 1.2$\,mag (to some
extent this could be due to extinction, which we cannot constrain),
the rise times are nearly identical, as can be seen in the top panel
of Figure \ref{f:05ci}.  The bottom panel shows a spectrum of
SN\,2005ci at $\sim 12$\,d after explosion compared to that of \cb\ at
a similar phase. The two spectra are qualitatively similar, although
the features of SN\,05ci are much weaker.  The velocities of the
hydrogen absorption minima are measured to be $\sim$ 15,000
km\,s$^{-1}$ and $\sim$ 13,000 km\,s$^{-1}$ for H$\alpha$ and
H$\beta$, respectively. These match well the velocities of
\cb\ at a comparable phase (Fig. \ref{velocities}), and they
are significantly higher than those measured for SNe\,1999em
and 1987A.

This additional identification raises the number of such SNe
discovered by the Lick Observatory SN search to two out of 52
SNe\,II-P discovered over the period studied by \citet{li11_LF}, in a
volume-limited sample. The corresponding rate is
$\sim 4^{+5.2}_{-2.6}$\% that of normal SNe\,II-P, where the dominant
error is derived from Poisson statistics. Since SNe\,II-P are the most
abundant SNe in the Universe, this is not an insignificant fraction,
on the order of 2\% of all core-collapse SNe. The rate we derive is
consistent with the one determined by \citet{smartt09} --- less than $\sim$
3\% --- by counting all SNe within about 28\,Mpc and assuming the sample
is reasonably complete and representative.

It would be of interest to study the metallicities of the locations of
these SNe, given the possibility that massive stars are more likely to
explode as BSGs than as RSGs at low metallicities
\citep[e.g.,][]{langer91,chieffi03}.  The location of \cb\ in the
outskirts of its host spiral galaxy suggests a relatively low
metallicity, in fact.  Recently, \citet{anderson10} have measured the
metallicity of H~II regions near the positions of CC~SNe, including
one in the general vicinity of SN\,2000cb, for which they find an oxygen
abundance ($8.70\pm0.13$ from the O3N2 method) that is close to typical for 
SNe\,II and II-P. However, that H~II region was more
than 3~kpc away from the supernova, so it may not give an accurate
estimate of the metallicity in the progenitor's immediate environment.

\section{Conclusion}

We have presented data and analysis of \cb, a peculiar Type II SN,
with an atypical light curve and extreme photospheric
velocities. \cb\ strongly resists our attempts to cast it into the
standard SN\,II-P category while also avoiding identity with
SN\,1987A. Photometrically it is intrinsically redder than SN\,1999em
and fainter by about 0.5\,mag on the plateau. It rises to maximum
brightness much more slowly than a normal SN\,II-P, yet this rise is
faster than that of SN\,1987A and proceeds with a distinctively
different colour and luminosity evolution. The main spectroscopic
peculiarity lies in the exceptionally high velocities of its features
at all times, most notably for the hydrogen absorption.

Our modeling, as well as extensive comparisons to other SNe, favour a
high-energy explosion of a relatively small-radius star, most probably
a BSG. We derive a rate for BSG explosions that is on the order of 2\%
of the rate of all core-collapse SNe.

SNe\,II-P are the most abundant SNe in the Universe and originate from
the explosion of RSGs, the most abundant evolved massive stars. SN\,1987A is
the keystone of modern Type II SN theory, despite being the explosion
of a BSG. Linking peculiar SNe such as SN\,1987A and \cb\ to regular
SNe\,II is critical for elucidating the processes that bring a star
from formation to obliteration.

\section*{Acknowledgments}

We thank Eddie Baron, Luc Dessart, Dan Maoz, and the referee (Stephen
Smartt) for useful discussions and comments on this manuscript; Mario
Hamuy and Andrea Pastorello for sharing their data; and the following
for assistance with observations: Zoltan Balog, Perry Berlind, Alison
Coil, Douglas Leonard, Maryam Modjaz, and Mark
Phillips. D.P. acknowledges support from an Einstein Fellowship and
from the US Department of Energy Scientific Discovery through Advanced
Computing (SciDAC) program under contract DE-FG02-06ER06-04.  A.V.F.'s
SN group at UC Berkeley is supported by National Science Foundation
(NSF) grant AST--0908886 and by the TABASGO Foundation. Supernova
research at the Harvard College Observatory is supported by NSF grant
AST--0907903.  The construction and ongoing operation of KAIT were
made possible by donations from Sun Microsystems, Inc., the
Hewlett-Packard Company, AutoScope Corporation, Lick Observatory, the
NSF, the University of California, the Sylvia \& Jim Katzman
Foundation, and the TABASGO Foundation. The Kast spectrograph at Lick
Observatory resulted from a generous donation made by Bill and Marina
Kast. We are grateful to the dedicated staffs at the Lick and
F. L. Whipple Observatories.

\vspace{.2cm}


	
	



\bibliographystyle{mn2e} 

\begin{thebibliography}{}

\bibitem[\protect\citeauthoryear{{Aldering} \& {Conley}}{{Aldering} \&
  {Conley}}{2000}]{aldering00}
{Aldering} G.,  {Conley} A.,  2000, \iaucirc, 7410, 3


\bibitem[Anderson et al.(2010)]{anderson10} Anderson, J.~P., 
Covarrubias, R.~A., James, P.~A., Hamuy, M., 
\& Habergham, S.~M.\ 2010, \mnras, 407, 2660 


\bibitem[Arcavi et al. (2009)]{arcavi09} Arcavi I., et al.,
2009, BAAS, 214, \#604.01

\bibitem[\protect\citeauthoryear{{Arnett}}{{Arnett}}{1979}]{arnett79}
{Arnett} W.~D.,  1979, \apj, 230, L37

\bibitem[\protect\citeauthoryear{{Arnett}, {Bahcall}, {Kirshner} \&
    {Woosley}}{{Arnett} et~al.}{1989}]{arnett89} {Arnett} W.~D.,
  {Bahcall} J.~N., {Kirshner} R.~P., {Woosley} S.~E., 1989, \araa, 27,
  629

\bibitem[\protect\citeauthoryear{{Baklanov}, {Blinnikov} \&
  {Pavlyuk}}{{Baklanov} et~al.}{2005}]{baklanov05}
{Baklanov} P.~V.,  {Blinnikov} S.~I.,  {Pavlyuk} N.~N.,  2005, Astronomy
  Letters, 31, 429

\bibitem[\protect\citeauthoryear{{Barbon}, {Benetti}, {Rosino},
    {Cappellaro} \& {Turatto}}{{Barbon} et~al.}{1990}]{barbon90}
  {Barbon} R., {Benetti} S., {Rosino} L., {Cappellaro} E., {Turatto}
  M., 1990, \aap, 237, 79

\bibitem[\protect\citeauthoryear{{Barbon}, {Ciatti} \&
    {Rosino}}{{Barbon} et~al.}{1979}]{barbon79} {Barbon} R., {Ciatti}
  F., {Rosino} L., 1979, \aap, 72, 287

\bibitem[Baron et al. (2000)]{baron00} Baron E., et al., 2000, \apj, 545, 444

\bibitem[\protect\citeauthoryear{{Bessell}}{{Bessell}}{1999}]{bessell99}
{Bessell} M.~S.,  1999, \pasp, 111, 1426

\bibitem[\protect\citeauthoryear{{Blinnikov}, {Lundqvist}, {Bartunov},
    {Nomoto} \& {Iwamoto}}{{Blinnikov} et~al.}{2000}]{blinnikov00}
  {Blinnikov} S., {Lundqvist} P., {Bartunov} O., {Nomoto} K.,
  {Iwamoto} K., 2000, \apj, 532, 1132

\bibitem[\protect\citeauthoryear{{Cardelli}, {Clayton} \&
    {Mathis}}{{Cardelli} et~al.}{1989}]{cardelli89} {Cardelli} J.~A.,
  {Clayton} G.~C., {Mathis} J.~S., 1989, \apj, 345, 245
  
  \bibitem[Chieffi et al.(2003)]{chieffi03} Chieffi, A., 
Dom{\'{\i}}nguez, I., H{\"o}flich, P., Limongi, M., 
\& Straniero, O.\ 2003, \mnras, 345, 111

\bibitem[\protect\citeauthoryear{{Dessart}, {Blondin}, {Brown}, {Hicken},
  {Hillier}, {Holland}, {Immler}, {Kirshner}, {Milne}, {Modjaz} \&
  {Roming}}{{Dessart} et~al.}{2008}]{dessart08} {Dessart} L., et al.,
  2008, \apj, 675, 644

\bibitem[Dessart \& Hillier (2008)]{dessart08b} Dessart L.,
  Hillier D.~J., 2008, \mnras, 383, 57

\bibitem[\protect\citeauthoryear{{Dessart} \& {Hillier}}{{Dessart} \& 
  {Hillier}}{2010a}]{dessart10a}
{Dessart} L., {Hillier} D.~J.,  2010a, \mnras, 405, 2141

\bibitem[Dessart et al. (2010b)]{dessart10b} Dessart L., Livne E.,  
   Waldman R., 2010b, \mnras, 408, 827


\bibitem[\protect\citeauthoryear{{Eastman}, {Woosley}, {Weaver} \&
    {Pinto}}{{Eastman} et~al.}{1994}]{eastman94} {Eastman} R.~G.,
  {Woosley} S.~E., {Weaver} T.~A., {Pinto} P.~A., 1994, \apj, 430, 300

\bibitem[\protect\citeauthoryear{{Fabricant}, {Cheimets}, {Caldwell}
    \& {Geary}}{{Fabricant} et~al.}{1998}]{fabricant98} {Fabricant}
  D., {Cheimets} P., {Caldwell} N., {Geary} J., 1998, \pasp, 110, 79

\bibitem[\protect\citeauthoryear{{Filippenko}}{{Filippenko}}{1982}]
  {filippenko82} {Filippenko} A.~V.,  1982, \pasp, 94, 715

\bibitem[\protect\citeauthoryear{{Filippenko}}{{Filippenko}}{1988}]
  {filippenko88} {Filippenko} A.~V.,  1988, \aj, 96, 1941

\bibitem[\protect\citeauthoryear{{Filippenko}}{{Filippenko}}{1997}]
  {filippenko97} {Filippenko} A.~V., 1997, \araa, 35, 309

\bibitem[\protect\citeauthoryear{{Filippenko}, {Li}, {Treffers} \&
    {Modjaz}}{{Filippenko} et~al.}{2001}]{filippenko01} {Filippenko}
  A.~V., {Li} W.~D., {Treffers} R.~R., {Modjaz} M., 2001, in
  {B.~Paczy\'{n}ski, W.-P.~Chen, \& C.~Lemme} ed., Small-Telescope 
  Astronomy on Global Scales (San Francisco: ASP, Vol.~246), 121

\bibitem[\protect\citeauthoryear{{Foley}, {Papenkova}, {Swift},
    {Filippenko}, {Li}, {Mazzali}, {Chornock}, {Leonard} \& {Van
      Dyk}}{{Foley} et~al.}{2003}]{foley03} {Foley} R.~J., et al.,
  2003, \pasp, 115, 1220

\bibitem[\protect\citeauthoryear{{Ganeshalingam}, {Li}, {Filippenko},
    {Anderson}, {Foster}, {Gates}, {Griffith}, {Grigsby}, {Joubert},
    {Leja}, {Lowe}, {Macomber}, {Pritchard}, {Thrasher} \&
    {Winslow}}{{Ganeshalingam} et~al.}{2010}]{ganeshalingam10}
  {Ganeshalingam} M., et al., 2010, \apjs, 190, 418

\bibitem[\protect\citeauthoryear{{Hamuy}}{{Hamuy}}{2003}]{hamuy03}
  {Hamuy} M., 2003, \apj, 582, 905

\bibitem[\protect\citeauthoryear{{Hamuy}}{{Hamuy}}{2004}]{hamuy04}
{Hamuy} M., 2004, in Measuring and Modeling the Universe, ed. W. L. 
Freedman (Pasadena: Carnegie Observatories) 

\bibitem[\protect\citeauthoryear{{Hamuy}, {Deng}, {Mazzali},
    {Morrell}, {Phillips}, {Roth}, {Gonzalez}, {Thomas-Osip},
    {Krzeminski}, {Contreras}, {Maza}, {Gonz{\'a}lez}}{{Hamuy} et~al.
    2009} {2009}]{hamuy09}
  {Hamuy} M., et al., 2009, \apj, 703, 1612

\bibitem[\protect\citeauthoryear{{Hamuy} \& {Pinto}}{{Hamuy} \&
    {Pinto}}{2002}]{hamuy02} {Hamuy} M., {Pinto} P.~A., 2002, \apj,
  566, L63

\bibitem[\protect\citeauthoryear{{Hamuy}, {Suntzeff}, {Gonzalez} \&
    {Martin}}{{Hamuy} et~al.}{1988}]{hamuy88} {Hamuy} M., {Suntzeff}
  N.~B., {Gonzalez} R., {Martin} G., 1988, \aj, 95, 63

\bibitem[\protect\citeauthoryear{{Hamuy}}{{Hamuy}}{2001}]{hamuy01b}
  {Hamuy} M.~A., 2001, PhD thesis, The University of Arizona

\bibitem[\protect\citeauthoryear{{Hanuschik} \& {Dachs}}{{Hanuschik}
    \& {Dachs}}{1987}]{hanuschik87} {Hanuschik} R.~W., {Dachs} J.,
  1987, \aap, 182, L29+

\bibitem[\protect\citeauthoryear{{Horne}}{{Horne}}{1986}]{horne86}
  {Horne} K., 1986, \pasp, 98, 609

\bibitem[\protect\citeauthoryear{{Jha}, {Challis}, {Kirshner} \&
    {Berlind}}{{Jha} et~al.}{2000}]{jha00} {Jha} S., {Challis} P.,
  {Kirshner} R., {Berlind} P., 2000, \iaucirc, 7410, 2

\bibitem[\protect\citeauthoryear{{Jones}, {Hamuy}, {Lira}, {Maza},
    {Clocchiatti}, {Phillips}, {Morrell}, {Roth}, {Suntzeff},
    {Matheson}, {Filippenko}, {Foley} \& {Leonard}}{{Jones}
    et~al.}{2009}]{jones09} {Jones} M.~I., et al., 2009, \apj, 696,
  1176

\bibitem[\protect\citeauthoryear{{Kasen} \& {Woosley}}{{Kasen} \&
    {Woosley}}{2009}]{kasen09} {Kasen} D., {Woosley} S.~E., 2009,
  \apj, 703, 2205

\bibitem[\protect\citeauthoryear{{Kirshner} \& {Kwan}}{{Kirshner} \&
    {Kwan}}{1974}]{kirshner74} {Kirshner} R.~P., {Kwan} J., 1974,
  \apj, 193, 27

\bibitem[\protect\citeauthoryear{{Kirshner}, {Oke}, {Penston} \&
    {Searle}}{{Kirshner} et~al.}{1973}]{kirshner73} {Kirshner} R.~P.,
  {Oke} J.~B., {Penston} M.~V., {Searle} L., 1973, \apj, 185, 303

\bibitem[Langer (1991)]{langer91} Langer N. 1991, \aap, 243, 155

\bibitem[\protect\citeauthoryear{{Leonard}, {Filippenko}, {Ardila} \&
    {Brotherton}}{{Leonard} et~al.}{2001}]{leonard01} {Leonard} D.~C.,
  {Filippenko} A.~V., {Ardila} D.~R., {Brotherton} M.~S., 2001, \apj,
  553, 861

\bibitem[\protect\citeauthoryear{{Leonard}, {Filippenko}, {Gates},
    {Li}, {Eastman}, {Barth}, {Bus}, {Chornock}, {Coil}, {Frink},
    {Grady}, {Harris}, {Malkan}, {Matheson}, {Quirrenbach} \&
    {Treffers}}{{Leonard} et~al.}{2002}]{leonard02a} {Leonard} D.~C.,
  et al., 2002a, \pasp, 114, 35 [Erratum: 114, 1291]

\bibitem[\protect\citeauthoryear{{Leonard}, {Filippenko}, {Li},
    {Matheson}, {Kirshner}, {Chornock}, {Van Dyk}, {Berlind},
    {Calkins}, {Challis}, {Garnavich}, {Jha} \& {Mahdavi}}{{Leonard}
    et~al.}{2002}]{leonard02b} {Leonard} D.~C., et al., 2002b, \aj,
  124, 2490

\bibitem[\protect\citeauthoryear{{Leonard}, {Kanbur}, {Ngeow} \&
    {Tanvir}}{{Leonard} et~al.}{2003}]{leonard03} {Leonard} D.~C.,
  {Kanbur} S.~M., {Ngeow} C.~C., {Tanvir} N.~R., 2003, \apj, 594, 247

\bibitem[\protect\citeauthoryear{{Li}, {Filippenko}, {Chornock} \&
    {Jha}}{{Li} et~al.}{2003}]{li03} {Li} W., {Filippenko} A.~V.,
  {Chornock} R., {Jha} S., 2003, \apj, 586, L9

\bibitem[\protect\citeauthoryear{{Li}, {Leaman}, {Chornock},
    {Filippenko}, {Poznanski}, {Ganeshalingam}, {Wang}, {Modjaz},
    {Jha}, {Foley} \& {Smith}}{{Li} et~al.}{2011}]{li11_LF} {Li} W.,
  et al., 2011, \mnras, in press (arXiv:1006.4612)

\bibitem[\protect\citeauthoryear{{Li}, {Filippenko}, {Treffers},
    {Friedman}, {Halderson}, {Johnson}, {King}, {Modjaz}, {Papenkova},
    {Sato} \& {Shefler}}{{Li} et~al.}{2000}]{li00} {Li} W.~D.,
    et al., 2000, in Cosmic Explosions, ed. {S.~S.~Holt \& W.~W.~Zhang}
    (New York: AIP), 103

\bibitem[\protect\citeauthoryear{{Madison} \& {Li}}{{Madison} \&
    {Li}}{2005}]{madison05} {Madison} D.~R., {Li} W., 2005, \iaucirc,
  8541, 1

\bibitem[\protect\citeauthoryear{{Matheson}, {Filippenko}, {Ho},
    {Barth} \& {Leonard}}{{Matheson} et~al.}{2000}]{matheson00}
  {Matheson} T., {Filippenko} A.~V., {Ho} L.~C., {Barth} A.~J.,
  {Leonard} D.~C., 2000, \aj, 120, 1499

\bibitem[\protect\citeauthoryear{{Matheson}, {Kirshner}, {Challis},
    {Jha}, {Garnavich}, {Berlind}, {Calkins}, {Blondin}, {Balog},
    {Bragg}, {Caldwell}, {Dendy Concannon}}{{Matheson} et~al.} {2008}]
    {matheson08} {Matheson} T., et al., 2008, \aj, 135, 1598

\bibitem[\protect\citeauthoryear{{Miller} \& {Stone}}{{Miller} \&
    {Stone}}{1993}]{miller93} {Miller} J.~S., {Stone} R.~P.~S., 1993,
  {Lick Obs. Tech. Rep. 66} (Santa Cruz: Lick Obs.)

\bibitem[\protect\citeauthoryear{{Modjaz}, {Kirshner}, {Challis} \&
    {Calkins}}{{Modjaz} et~al.}{2005}]{modjaz05} {Modjaz} M.,
  {Kirshner} R., {Challis} P., {Calkins} M., 2005, \iaucirc, 8542, 2

\bibitem[\protect\citeauthoryear{{Nadyozhin}}{{Nadyozhin}}{2003}]{nadyozhin03}
  {Nadyozhin} D.~K., 2003, \mnras, 346, 97

\bibitem[\protect\citeauthoryear{{Nugent}, {Sullivan}, {Ellis},
    {Gal-Yam}, {Leonard}, {Howell}, {Astier}, {Carlberg}, {Conley},
    {Fabbro}, {Fouchez}, {Neill}, {Pain}, {Perrett}, {Pritchet} \&
    {Regnault}}{{Nugent} et~al.}{2006}]{nugent06} {Nugent} P., et al.,
  2006, \apj, 645, 841

\bibitem[\protect\citeauthoryear{{Olivares}}{{Olivares}}{2010}]{olivares10}
{Olivares} F., et al.,  2010, \apj, 715, 833 

\bibitem[\protect\citeauthoryear{{Papenkova} \& {Li}}{{Papenkova} \&
    {Li}}{2000}]{papenkova00} {Papenkova} M., {Li} W.~D., 2000,
  \iaucirc, 7410, 1

\bibitem[\protect\citeauthoryear{{Pastorello}, {Baron}, {Branch},
    {Zampieri}, {Turatto}, {Ramina}, {Benetti}, {Cappellaro}, {Salvo},
    {Patat}, {Piemonte}, {Sollerman}, {Leibundgut} \&
    {Altavilla}}{{Pastorello} et~al.}{2005}]{pastorello05}
  {Pastorello} A., et al., 2005, \mnras, 360, 950

\bibitem[\protect\citeauthoryear{{Popov}}{{Popov}}{1993}]{popov93}
  {Popov} D.~V., 1993, \apj, 414, 712

\bibitem[\protect\citeauthoryear{{Poznanski}, {Butler}, {Filippenko},
    {Ganeshalingam}, {Li}, {Bloom}, {Chornock}, {Foley}, {Nugent},
    {Silverman}, {Cenko}, {Gates}, {Leonard}, {Miller}, {Modjaz},
    {Serduke}, {Smith}, {Swift} \& {Wong}}{{Poznanski}
    et~al.}{2009}]{poznanski09} {Poznanski} D., et al., 2009, \apj,
  694, 1067

\bibitem[\protect\citeauthoryear{{Poznanski}, {Nugent} \&
    {Filippenko}}{{Poznanski} et~al.}{2010}]{poznanski10} Poznanski,
  D., Nugent, P.~E., \& Filippenko, A.~V.\ 2010, \apj, 721, 956

\bibitem[\protect\citeauthoryear{{Richmond}, {Treffers}, {Filippenko},
    {Paik}, {Leibundgut}, {Schulman} \& {Cox}}{{Richmond}
    et~al.}{1994}]{richmond94} {Richmond} M.~W., {Treffers} R.~R.,
  {Filippenko} A.~V., {Paik} Y., {Leibundgut} B., {Schulman} E., {Cox}
  C.~V., 1994, \aj, 107, 1022

\bibitem[\protect\citeauthoryear{{Schlegel}, {Finkbeiner} \&
    {Davis}}{{Schlegel} et~al.}{1998}]{schlegel98} {Schlegel} D.~J.,
  {Finkbeiner} D.~P., {Davis} M., 1998, \apj, 500, 525

\bibitem[\protect\citeauthoryear{{Schmitz} \& {Gaskell}}{{Schmitz} \&
    {Gaskell}}{1988}]{schmitz88} {Schmitz} M.~F., {Gaskell} C.~M.,
  1988, in Supernova 1987A in the Large Magellanic Cloud,
  ed. {M.~Kafatos \& A.~G.~Michalitsianos} (Cambridge: Cambridge
  Univ. Press), 112

\bibitem[\protect\citeauthoryear{{Shigeyama} \& {Nomoto}}{{Shigeyama}
    \& {Nomoto}}{1990}]{shigeyama90} {Shigeyama} T., {Nomoto} K.,
  1990, \apj, 360, 242


\bibitem[\protect\citeauthoryear{{Smartt}, {Eldridge}, {Crockett} \&
    {Maund}}{{Smartt} et~al.}{2009}]{smartt09} {Smartt} S.~J.,
  {Eldridge} J.~J., {Crockett} R.~M., {Maund} J.~R., 2009, \mnras,
  395, 1409

\bibitem[\protect\citeauthoryear{{Springob}, {Masters}, {Haynes},
    {Giovanelli} \& {Marinoni}}{{Springob} et~al.}{2009}]{springob09}
  {Springob} C.~M., {Masters} K.~L., {Haynes} M.~P., {Giovanelli} R.,
  {Marinoni} C., 2009, \apjs, 182, 474

\bibitem[\protect\citeauthoryear{{Storm}, {Carney}, {Gieren},
    {Fouqu{\'e}}, {Latham} \& {Fry}}{{Storm} et~al.}{2004}]{storm04}
  {Storm} J., {Carney} B.~W., {Gieren} W.~P., {Fouqu{\'e}} P.,
  {Latham} D.~W., {Fry} A.~M., 2004, \aap, 415, 531

\bibitem[\protect\citeauthoryear{{Turatto}, {Benetti} \&
    {Cappellaro}}{{Turatto} et~al.}{2003}]{turatto03} {Turatto} M.,
  {Benetti} S., {Cappellaro} E., 2003, in From Twilight to Highlight:
  The Physics of Supernovae, ed. {W.~Hillebrandt \& B.~Leibundgut}
  (Berlin: Springer-Verlag), 200

\bibitem[\protect\citeauthoryear{{Utrobin} \& {Chugai}}{{Utrobin} \&
    {Chugai}}{2009}]{utrobin09} {Utrobin} V.~P., {Chugai} N.~N., 2009,
  \aap, 506, 829

\bibitem[\protect\citeauthoryear{{Utrobin}, {Chugai} \&
    {Andronova}}{{Utrobin} et~al.}{1995}]{utrobin95} {Utrobin} V.~P.,
  {Chugai} N.~N., {Andronova} A.~A., 1995, \aap, 295, 129

\bibitem[\protect\citeauthoryear{{Utrobin}, {Chugai} \&
    {Pastorello}}{{Utrobin} et~al.}{2007}]{utrobin07} {Utrobin} V.~P.,
  {Chugai} N.~N., {Pastorello} A., 2007, \aap, 475, 973

\bibitem[\protect\citeauthoryear{{Wade} \& {Horne}}{{Wade} \&
    {Horne}}{1988}]{wade88} {Wade} R.~A., {Horne} K., 1988, \apj, 324,
  411

\bibitem[\protect\citeauthoryear{{Woodings}, {Williams}, {Martin},
    {Burman} \& {Blair}}{{Woodings} et~al.}{1998}]{woodings98}
  {Woodings} S.~J., {Williams} A.~J., {Martin} R., {Burman} R.~R.,
  {Blair} D.~G., 1998, \mnras, 301, L5

\bibitem[\protect\citeauthoryear{{Woosley}}{{Woosley}}{1988}]{woosley88}
  {Woosley} S.~E., 1988, \apj, 330, 218

\bibitem[\protect\citeauthoryear{{Woosley}, {Pinto}, {Martin} \&
    {Weaver}}{{Woosley} et~al.}{1987}]{woosley87} {Woosley} S.~E.,
  {Pinto} P.~A., {Martin} P.~G., {Weaver} T.~A., 1987, \apj, 318, 664

\bibitem[\protect\citeauthoryear{{Woosley} \& {Weaver}}{{Woosley} \&
    {Weaver}}{1995}]{woosley95} {Woosley} S.~E., {Weaver} T.~A., 1995,
  \apjs, 101, 181

\bibitem[\protect\citeauthoryear{{Young}}{{Young}}{2004}]{young04}
  {Young} T.~R., 2004, \apj, 617, 1233

\end{thebibliography}

\onecolumn

\begin{deluxetable}{cccccc}
\tablewidth{0pt}
\tabletypesize{\scriptsize}
\tablecaption{Unfiltered and {\it BVI} photometry of SN\,2000cb.\label{phot_table}}
\tablehead{
\colhead{JD} &
\colhead{Phase\tablenotemark{a}} &
\colhead{$Clear$ (mag)}&
\colhead{$B$ (mag)} &
\colhead{$V$ (mag)}&
\colhead{$I$ (mag)}}
\startdata
2451658.96 &   3 & 18.72(06) & \ldots    & \ldots    & \ldots    \\ 
2451661.97 &   6 & 17.74(06) & \ldots    & \ldots    & \ldots    \\ 
2451663.92 &   8 & 17.83(06) & 18.78(06) & 18.04(03) & 17.66(06) \\ 
2451665.96 &  10 & 17.42(06) & \ldots    & \ldots    & \ldots    \\ 
2451667.93 &  12 & \ldots    & 18.78(07) & 17.80(04) & 17.30(06) \\ 
2451670.96 &  15 & 17.06(06) & \ldots    & \ldots    & \ldots    \\ 
2451691.93 &  36 & 16.33(06) & \ldots    & \ldots    & \ldots    \\ 
2451692.88 &  37 & \ldots    & 17.73(03) & 16.62(02) & 15.87(02) \\ 
2451693.93 &  38 & 16.27(06) & \ldots    & \ldots    & \ldots    \\ 
2451696.87 &  41 & 16.11(06) & 17.83(05) & 16.61(02) & 15.80(03) \\ 
2451699.88 &  44 & 16.11(06) & \ldots    & \ldots    & \ldots    \\ 
2451700.85 &  45 & \ldots    & 17.87(02) & 16.56(02) & 15.75(02) \\ 
2451702.87 &  47 & 16.24(06) & \ldots    & \ldots    & \ldots    \\ 
2451706.85 &  51 & 16.14(06) & 17.98(05) & 16.59(02) & 15.67(01) \\ 
2451713.82 &  58 & \ldots    & 18.05(04) & 16.61(02) & 15.63(02) \\ 
2451716.85 &  61 & 16.04(06) & \ldots    & \ldots    & \ldots    \\ 
2451717.84 &  62 & \ldots    & 18.08(04) & 16.56(02) & 15.59(02) \\ 
2451719.81 &  64 & 16.19(06) & \ldots    & \ldots    & \ldots    \\ 
2451721.82 &  66 & \ldots    & 18.14(04) & 16.56(01) & 15.57(02) \\ 
2451722.84 &  67 & 16.18(06) & \ldots    & \ldots    & \ldots    \\ 
2451726.79 &  71 & 16.11(06) & \ldots    & \ldots    & \ldots    \\ 
2451728.77 &  73 & \ldots    & 18.19(05) & 16.57(02) & 15.58(01) \\ 
2451731.78 &  76 & 16.10(06) & \ldots    & \ldots    & \ldots    \\ 
2451735.72 &  80 & \ldots    & 18.20(04) & 16.63(02) & 15.61(02) \\ 
2451740.75 &  85 & 16.20(06) & \ldots    & \ldots    & \ldots    \\ 
2451742.73 &  87 & \ldots    & 18.38(06) & 16.66(02) & 15.66(02) \\ 
2451743.78 &  88 & 16.23(13) & \ldots    & \ldots    & \ldots    \\ 
2451747.72 &  92 & 16.21(06) & \ldots    & \ldots    & \ldots    \\ 
2451749.69 &  94 & \ldots    & 18.55(06) & 16.85(02) & 15.74(02) \\ 
2451752.72 &  97 & 16.34(06) & \ldots    & \ldots    & \ldots    \\ 
2451756.69 & 101 & 16.61(06) & 18.78(09) & 17.18(02) & 15.93(02) \\ 
2451760.68 & 105 & 16.90(06) & \ldots    & \ldots    & \ldots    \\ 
2451770.69 & 115 & 16.98(06) & \ldots    & \ldots    & \ldots    \\ 
2451774.66 & 119 & 16.89(07) & \ldots    & \ldots    & \ldots    \\ 
2451781.66 & 126 & \ldots    & 19.37(16) & 17.90(05) & 16.52(03) \\ 
2451788.66 & 133 & \ldots    & 19.37(18) & 18.14(19) & \ldots    \\ 
2451795.64 & 140 & \ldots    & 19.32(20) & 18.20(10) & 16.68(03) \\ 
2451802.64 & 147 & \ldots    & \ldots    & 18.21(15) & 16.80(07) \\
\enddata 
\tablecomments{Values in parentheses are 1$\sigma$ measurement uncertainties in
  hundredths of mag.}
\tablenotetext{a}{Relative to JD $=2451656$.}
\end{deluxetable}

\begin{deluxetable}{ccccccc}
\tablewidth{0pt}
\tabletypesize{\scriptsize}
\tablecaption{Comparison stars for the field of SN\,2000cb.\label{stand_table}}
\tablehead{
\colhead{Star} &
\colhead{$\alpha_{\rm J2000}$} &
\colhead{$\delta_{\rm J2000}$ }&
\colhead{$B$ (mag)}&
\colhead{$V$ (mag)}&
\colhead{$I$ (mag)}&
\colhead{$N_{\rm calib}\tablenotemark{a}$}}
\startdata
1 & 16:01:24.72 &   +1:45:12.95  &  18.454 (005)  & 16.948 (007)  &  15.039 (004)  & 3\tablenotemark{b} \\
2 & 16:01:37.57 &   +1:45:02.35  &  16.034 (010)  & 15.355 (011)  &  14.562 (011) &  4 \\ 
3 & 16:01:29.31 &   +1:44:54.15 &  17.965 (009)  & 17.269 (014)  &  16.468 (013) &  4\tablenotemark{b} \\ 
4 & 16:01:37.68 &   +1:44:01.86 &  16.058 (011)  & 15.245 (010)  &  14.370 (005)  & 4 \\ 
5 & 16:01:42.60 &   +1:43:54.82 &  15.331 (012)  & 14.637 (009)  &  13.873 (011)  & 4 \\
6 & 16:01:37.97 &   +1:43:15.13 &  16.444 (015)  & 15.609 (013)  &  14.721 (004)  & 4 \\
7 & 16:01:40.84 &   +1:43:09.37 &  16.944 (009)  & 16.305 (008)  &  15.547 (006)  & 4 \\ 
8 & 16:01:41.42 &   +1:42:49.59 &  17.800 (012)  & 17.156 (003)  &  16.482 (007)  & 4\tablenotemark{b} \\
9 & 16:01:25.05 &   +1:42:36.87 &  16.987 (013)  & 16.394 (012)  &  15.638 (004)  & 3\tablenotemark{c} \\
10 & 16:01:26.98 &   +1:42:14.91 &  17.544 (014)  & 16.849 (008)  &  16.039 (011)  & 4 \\ 
11 & 16:01:25.51 &   +1:42:29.21 &  18.994 (006)  & 17.610 (015)  &  15.882 (012)  & 3\tablenotemark{b} \\
12 & 16:01:35.25 &  +1:41:02.21 &  16.905 (009)  & 16.229 (007)  &  15.420 (003)  & 4 \\ 
13 & 16:01:22.75 &   +1:40:58.56 &  18.695 (030)  & 17.391 (000) & 15.837 (001)  & 2\tablenotemark{b} \\
14 & 16:01:34.33 &   +1:39:49.91 &  18.143 (016)  & 17.348 (017)  &  16.321 (013)  & 2 \\
15 & 16:01:34.61 &   +1:45:25.55 &  15.842 (012)  & 14.952 (003)  &  14.006 (012)  & 4 \\
\enddata 
\tablecomments{1$\sigma$ uncertainties are in units of $10^{-3}$ mag.}
\tablenotetext{a}{Number of calibration epochs.}
\tablenotetext{b}{One calibration fewer in the $V$ band.}
\tablenotetext{c}{One calibration fewer in the $I$ band.}
\end{deluxetable}

\begin{deluxetable}{llcccll}
\tablewidth{0pt}
\tabletypesize{\scriptsize}
\tablecaption{Journal of spectroscopic observations of SN\,2000cb.\label{spec_table}}
\tablehead{
\colhead{UT Date} &
\colhead{Phase (d)} &
\colhead{Range (\AA)}&
\colhead{Exp (s)}&
\colhead{Airmass\tablenotemark{a}}&
\colhead{Observer(s)\tablenotemark{b}}&
\colhead{Instrument}}
\startdata
2000 April 28.37 & 6.9 & 3720--7540  & $2 \times 1200$ & 1.2  & PB & FAST\\
2000 April 30.38 & 8.9 & 3720--7540 & 1200 & 1.2  & ZB & FAST\\
2000 May 31.372 & 39.9 & 4250--6950 & 900 & 1.3  & AF, AC & Kast\\ 
2000 Jun. 27.265 & 66.8 & 4240--6950 & 1200 & 1.3  & AF, AC & Kast\\
2000 Jul. 6.377 & 75.9 & 4250--6960 & 1200 & 2.5 & AF, RC, MM  & Kast\\
2000 Jul. 28.308 & 97.8 & 4250--6940 & 900 & 2.2  & AF, RC & Kast\\ 
2000 Aug. 26.198 & 126.7 & 4240--6950 & 1500 & 1.8  & DL, WL & Kast\\ 
2000 Sep. 6.216 & 137.7 & 3300--7760 & 1800 & 2.8  & AF, WL, RC & Kast\\
2000 Sep. 26.136 & 157.6 & 3300--7800 & 1800 & 2.2 & AF, WL, RC, MP & Kast \\
\enddata 
\tablecomments{Spectra were obtained using the Kast spectrograph on
  the 3\,m Shane telescope at Lick Observatory and the FAST
  spectrograph on the 1.5\,m Tillinghast telescope at Whipple
  Observatory.}
  \tablenotetext{a}{Airmass corresponds to the middle of the exposure.}
\tablenotetext{b}{AC = Alison Coil, AF = Alex Filippenko, DL = Douglas
  Leonard, MM = Maryam Modjaz, MP = Mark Phillips, PB = Perry Berlind,
  RC = Ryan Chornock, WL = Weidong Li, ZB = Zoltan Balog.}
\end{deluxetable}

\begin{deluxetable}{ccc}
\tablewidth{0pt}
\tabletypesize{\scriptsize}
\tablecaption{Unfiltered photometry of SN\,2005ci.\label{phot05ci_table}}
\tablehead{
\colhead{JD} &
\colhead{Phase\tablenotemark{a}} &
\colhead{$Clear$ (mag)}}
\startdata
2453523.86 & 3 & 20.16(15)  \\
2453531.81 & 11 & 18.78(13)  \\
2453532.85 & 12 & 18.75(13)  \\
2453533.83 & 13 & 18.76(13)  \\
2453534.76 & 14 & 18.86(13)  \\
2453536.80 & 16 & 18.51(16)  \\
2453540.82 & 20 & 18.31(13)  \\
2453545.79 & 25 & 18.18(13)  \\
2453549.78 & 29 & 17.94(13)  \\
2453553.79 & 33 & 17.75(13)  \\
2453557.78 & 37 & 17.63(13)  \\
2453561.76 & 41 & 17.72(15)  \\
2453565.75 & 45 & 17.53(13)  \\
2453567.68 & 47 & 17.68(15)  \\
\enddata 
\tablecomments{Values in parentheses are 1$\sigma$ measurement uncertainties in
  hundredths of mag.}
\tablenotetext{a}{Relative to JD $=2453521$.}
\end{deluxetable}

\end{document}